# A REVIEW OF THE ANOMALIES IN DIRECTED ENERGY DEPOSITION (DED) PROCESSES AND POTENTIAL SOLUTIONS

*Michael Liu, Abhishek Kumar, Satish Bukkapatnam, Mathew Kuttolamadom*
*Texas A&M University, College Station, TX, USA*

**Abstract**

Directed Energy Deposition (DED) processes allow additive manufacturing and repair of metallic components with generatively-designed complex geometries, and excellent compositional control. However, its applicability and adoption has been limited when compared to powder bed fusion (PBF) because several issues and anomalies innate to the process are yet to be suitably understood and resolved. This work catalogs and delineates these anomalies in the DED process along with their causes and solutions, based on a state-of-the-art literature review. This work also serves to enumerate and associate the underlying causes to the detrimental effects which manifest as undesirable part/process outcomes. These DED-specific anomalies are categorized under groups related to the part, process, material, productivity, safety, repair, and composition. Altogether, this primer acts as a guide to best prepare for and mitigate the problems that are encountered in DED, and also to lay the groundwork to inspire novel solutions to further advance DED into mainstream manufacturing.

## 1. INTRODUCTION

Many Additive Manufacturing (AM) technologies trace their origins to the innovations made during this past century in precursor utilization and material consolidation for powder metallurgy (P/M), laser welding, and arc processes [1]. Metal AM was first developed in the 1930's through manual layer-by-layer builds created by arc-welding and high pressure [2]. A new process would be created afterward wherein molten beads of weld material could be layered in succession to fuse and create a bottom up structure [3]. The evolution of 'welding-based AM' would begin during the 1960s and 1970s where patents on rotary-based, continual arc-welding for AM of solid structures and more were filed [4-6]. During the 1980s, a paradigm shift in AM metal working began with patents from Housholder and Brown which initiated the infancy of laser based AM [7, 8]. Housholder developed a powder based feed technology while Brown created the foundation for directed energy deposition (DED), which was further improved via the technological innovations of Lewis and Jeantette [9, 10].

In the last two decades, AM has proliferated beyond the confines of scientific research into prototyping and production. Among AM processes, DED, often accomplished via techniques such as Laser Engineered Net Shaping (LENS) is eliciting particular interest for adding material to and repairing existing parts, besides bulk part manufacture. It employs a coaxial laser beam along with multiple powder-feeding nozzles, which in turn spot-deposits and fuses material at the desired location, as the deposition assembly moves via Computer Numerical Control (CNC). Further, the flexibility of the powder delivery system allows for the development of functionally-graded parts with variable material compositions, and hence location-dependent properties, besides repair and coating applications of parts [11-16].

The feasibility of using production-grade AM parts requires precise control of the process to impart desired functionality. Understanding of the process mechanics and defect pathways, crucial for effective process control, is rather limited for the LENS-DED process. As in other AM processes, there are myriad of factors (Figure 1) on which the final build quality depends. These factors contribute combinatorially to the part defects and other anomalies in the LENS-DED process [1]. Such a delineation of the factors at play is important towards mitigating the part defects and address the safety and productivity issues that manifest. The parameters above the central rib of the fish-bone in Figure 1 are pre-defined (also referred to as 'constant' factors in the experimental design literature) and latent (state) variables. Many of these may not be monitored and are cumbersome to vary during the process. These include the properties of the stock material used in the process and machine specifications, among others. Certain other pre-defined factors that determine the initial conditions of the process environment and the real-time process state that need to be monitored; these include the chamber conditions, gas flow, temperature, etc. The ones below the rib line are controllable and can be rectified as a remedial compensation during or before the process. They can be modified as desired. Real-time monitoring and adjustments of these factors are also supported. These include process variables like laser power, feed rate, etc. Among these, an earlier study [1] suggests that the most common vital parameters are scanning speed, laser power, powder feed rate, hatch spacing, interlayer dwell time, laser beam diameter, and laser scanning pattern [1].

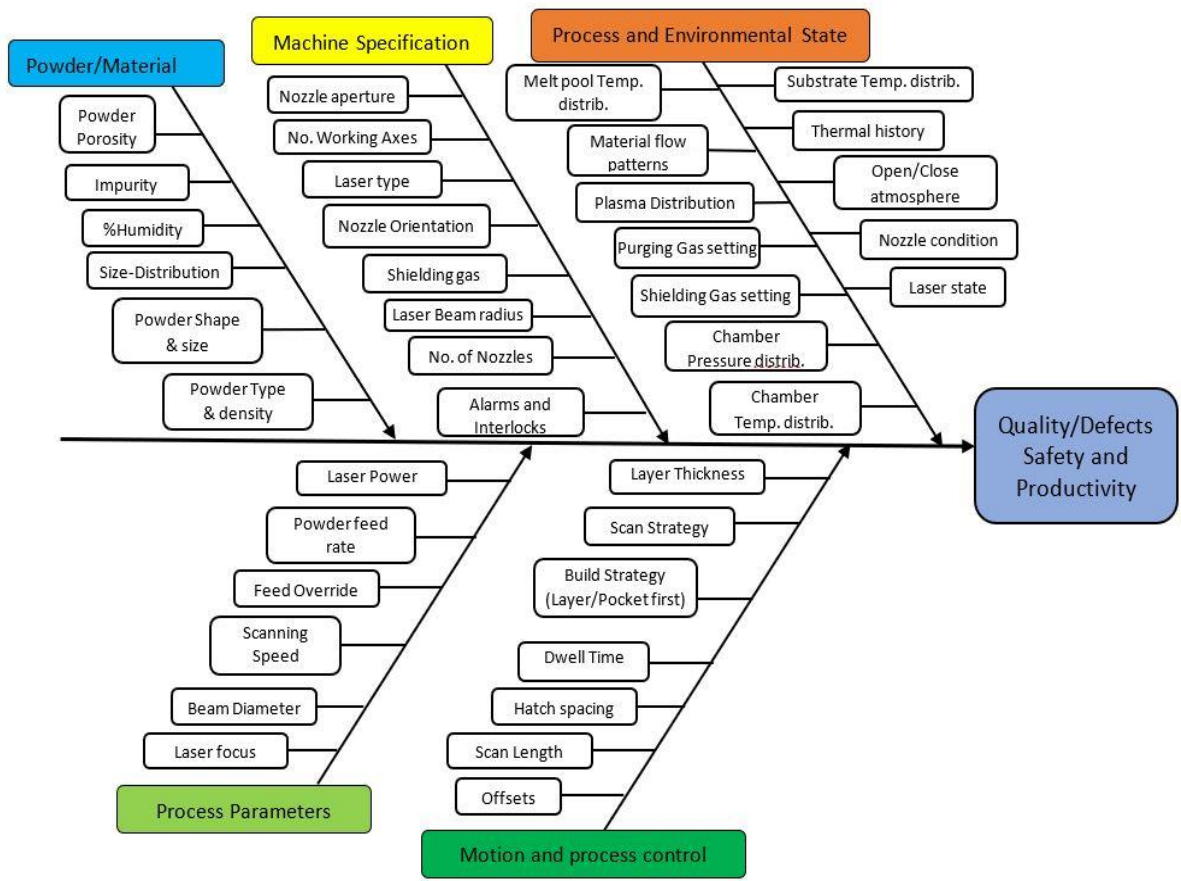

*Figure 1 : A representative list of material/process parameters and control variables that influence the quality, safety and productivity of the LENS-DED process.*

Additionally, a process parameter or a factor can contribute to multiple part defects, and impair the process efficiency and safety (see Figure 2). The depicted multivariate relation can be used for a root-cause analysis to remedy the process to a more desirable state. To mitigate a defect, the corresponding parameters of influence should be monitored to gain a better understanding of process stochastics. For example, when warping manifests in the part, the heat map guides us to optimize the laser power, scan length and dwell time, thus enabling one to address a defect or anomaly. Some of these issues could be further corrected by post-processing, albeit at the cost of productivity. However, the recipe for success and productivity takes root in the pre- and in-process operations to a great extent.

**Pre-Defined:**

| Powder type | Composition | Density | | | | | |
|---|---|---|---|---|---|---|---|
| Powder Shape | Flowability | Density | | | | | |
| Powder Size | Flowability | Density | | | | | |
| Size Distribution | Dimensional tolerance | Density | Morphology | Production Rate | | | |
| Powder Porosity | Density | Porosity | | | | | |

**Latent process state and control variables:**

| Laser power | Shrinkage | Balling | Warping | Waviness | Production Rate | | |
|---|---|---|---|---|---|---|---|
| Hatch Spacing | Shrinkage | Porosity | Surface Roughness | Anisotropy | Morphology | Melt pool size | Waviness |
| | Dimensional tolerance | Production Rate | Energy Density | Surface Deformation | | | |
| Scan Speed | Shrinkage | Porosity | Energy Density | Surface Roughness | Morphology | Melt pool size | Balling |
| Layer Thickness | Shrinkage | Porosity | Energy Density | Density | Bond Strength | Production Rate | Staircase Effect |
| Gas flow (rate and direction) | Dimensional tolerance | Porosity | Bond Strength | Anisotropy | | | |
| Scan Pattern | Surface Roughness | Porosity | Density | Anisotropy | | | |
| Scan Length | Shrinkage | Warping | Fracture | | | | |
| Substrate Temperature | Balling | Surface Deformation | Melt pool size | | | | |
| Build Direction | Dimensional tolerance | Anisotropy | Bond Strength | | | | |
| Dwell Time | Shrinkage | Warping | Dimensional tolerance | Morphology | Surface Deformation | | |

*Figure 2: Heat map of LENS-DED anomalies and the effect of process parameters.*

Due to the complexity of the LENS-DED process, understanding and tracking of process anomalies can facilitate a streamlined approach to address quality, productivity, and safety issues. The work here offers issues surrounding LENS-DED along with current and proposed solutions to mitigate them. Many anomalies addressed here have been encountered by the authors during their investigations on Co-Cr-Mo [17], Stainless Steel 316L [] and other materials. The motivation for this work stems from the difficulties faced during manufacture and the benefit that operators could obtain by referencing the issues raised here. Similar to online compendiums that offer a cache of insights to various problems, this work offers an anomaly-based reference, grouped under intuitive categories. Previous studies attempted to classify metallurgical defects in parts developed by DED into two types: process-induced defects and material-induced defects [18]. A delineation of the process anomalies that can have implications beyond part defects does not exist. This assessment uses a more comprehensive but connected approach to help diagnose and catalog anomalies.

The paper is organized as follows. An overview of LENS-DED-related anomalies when utilizing metal powders and their grouping is presented in Section 2. Sections 3-9 present a review of each of the seven categories of anomalies, and possible solutions for their mitigation. Section 10 discusses the key implications of these anomalies on the DED process and presents an overview of future directions. Note that the present discussion is limited to laser based DED of metals using powder stock (LENS-DED); every reference to a DED process in the remainder of this document specifically refers to LENS-DED.

## 2. CLASSIFICATION OF DED ANOMALIES AND SCOPE OF STUDY

As summarized in Figure 3, we classify DED anomalies into seven major categories, namely, part quality and defects, process and machine tool faults, material issues, productivity issues, safety issues, repair-specific issues, and compositional issues. It is plausible that some of the issues discussed might fit in multiple categories, but it is placed in the most relevant categories.

1. **Part quality and defects:** The manifestation of these issues can be found on the AM parts by probing over multiple spatial and spectral scales. DED defects that exist as markers in the part or within the microstructure are important to realize to improve the quality of end use parts. These in turn can be further classified into three sub-categories.
    a) **Geometrical:** These issues are manifested as deviation from nominal geometry (perfect) and its allowable variation (tolerance). These issues lead to variation in form and dimension of individual features from as-modeled or as-intended geometry. Employing strategies to mitigate them reduces the need for post-processing and hence the machine running time.
    b) **Morphological:** This class of defects affects the surface texture, surface finish, and surface topography, as well as the bulk defects, such as pores, cavities, unfused powders and bulk inclusions.
    c) **Microstructural:** These defects are related to microstructural aspects of the surface and sub-surface. Any irregularities having to do with aesthetic quality of the part such as discoloration or surface defects would also pertain to this group.

2. **Process and machine tool faults:** The anomalies arise by the virtue of erroneous process parameters and/or working conditions. Process and machine tool faults are often caused by anomalous parameters and machine settings. As DED 'delivers' laser power and material powder simultaneously, several considerations must be met to successfully deposit material upon a substrate and ensuing layered structures. The powder delivery system of DED provides design freedom without the need of supports; however, the flow characteristics of powder affect the structural integrity of the part. DED also harbors issues related to the laser including inefficiency of power, spattering, lack of complete melting and fusion, and laser attenuation caused by artifacts of the build and precursor materials. These process faults impart product defects, and can lead to safety and productivity issues. These in turn are classified into two sub-categories:

    a) **Anomalous machine or process settings:** These anomalies arise due to faulty machine settings. These may also arise on account of erroneous gauges readings as well as exogenous factors contributing to undesirable internal (latent) process states.

    b) **Controller Error:** These anomalies arise on account of communication and actuation errors in the controller. Especially in the DED process of these errors may arise due to malicious cybersecurity attacks [MaheshProcIEEE20]. These anomalies may adversely affect not just the product quality, but also the safety and productivity in manufacturing.

3. **Stock material issues:** These issues arise due to the inherent defects and variations in the raw material, *i.e.*, the powder or improper handling of the powder stock. These issues can include inconsistencies and defects in the stock material which are manifested in the part as defects and other anomalies. Defect free and consistent stock material favor consistent and cohesive deposit and microstructure. Impurities in the stock powder can implant itself in the part and impede the mechanical integrity. These issues can be rectified by examining and pre-processing the powder used to manufacture the component.

4. **Productivity issues:** Productivity issues address those which impact the machine running time and add to the manufacturing cost. Moreover, these issues can impede the manufacturing feasibility and render the process inefficient as compared to the conventional counterparts. For bringing the process to mainstream manufacturing, the production cost must be minimized to bring it on par with comparable traditional manufacturing methods. These issues toll on the total cost of the process and might make the processing cost prohibitive to produce the component additively. These issues also affect the rate of production.

5. **Safety issues:** These anomalies have to do with the machine itself that may endanger the operator or the machine during the build. The stock powder must be dealt with utmost éclat to avoid the combustion and the high energy laser can pose serious threat to the operator. The most pressing of these concerns are catastrophic hazards such as those associated to fire safety and crashing the nozzle into the sample. Less detrimental issues are also presented such as contamination due to oil leaking from LENS machines with subtractive components. These issues pose a threat to the operator or anyone who is exposed to the process. The issues need be dealt meticulously to avert accidents and adverse effects on the environment.

6. **Repair-specific issues**: The DED process is considered in the industry for various component repair applications. A few issues pertaining to the efficiency and quality of the original/rebuilt component hamper the application of DED for repair applications.

7. **Compositional issues:** DED allows multiple material powders to flow out of a nozzle, each at distinct and independently tunable rates into the melt pool (i.e., the laser deposition region). This facilitates achieve compositional gradients necessary to create functionally graded material structure. Certain issues arise specifically during DED-printing of functionally-graded materials and structures.

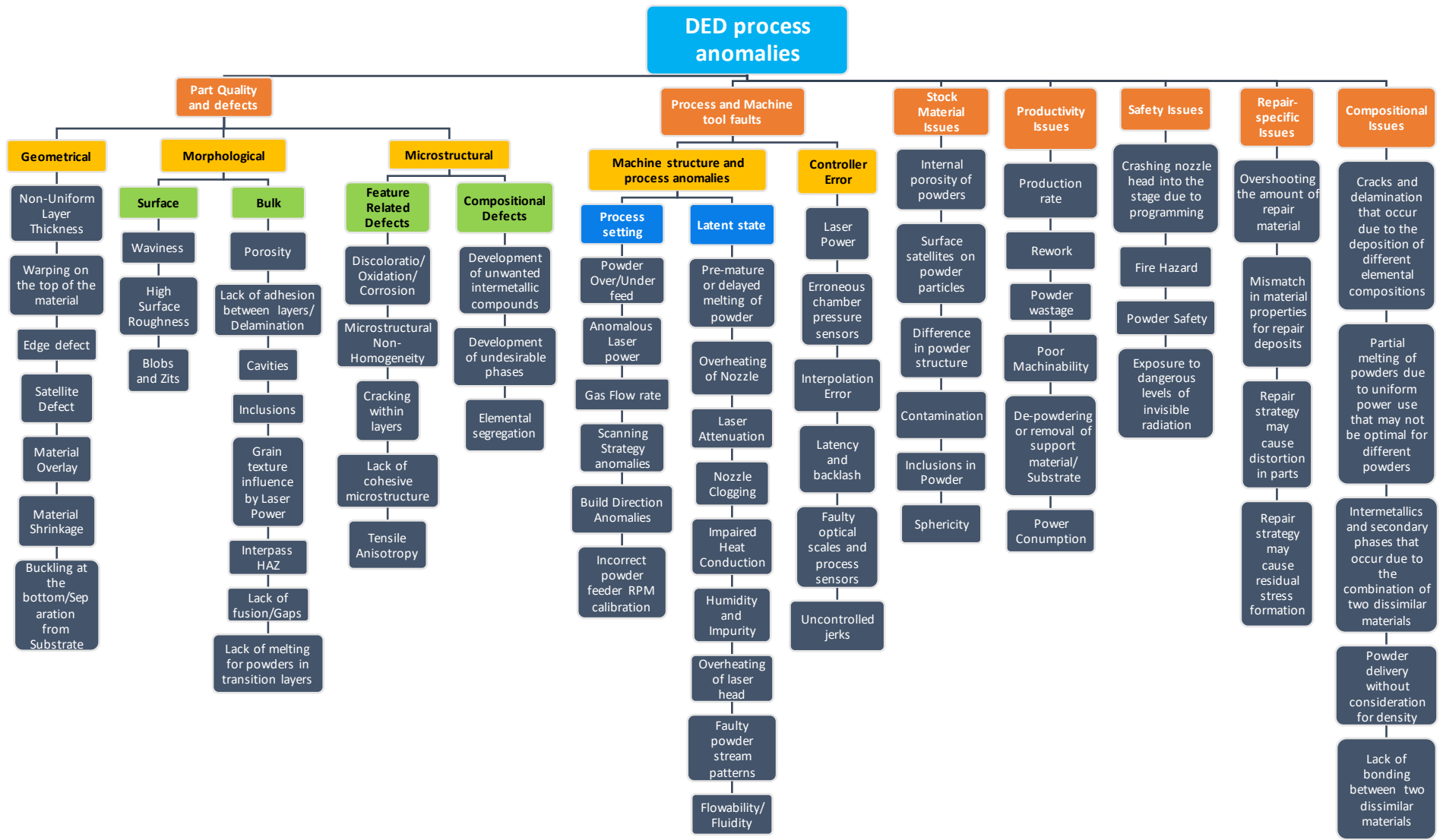

*Figure 3: Flow chart of process anomaly classification.*

## 3. PART QUALITY AND DEFECTS
These issues affect the final build quality and the structural integrity of the part produced by DED. These defects are further resolved into (i) Geometrical, (ii) Morphological and (iii) Microstructural. Each type can be analyzed by probing the part at macro- and/or micro-scales.

### 3.1 Geometrical Defects:
The deviation of the desired contour as modeled in CAD from the produced part is grouped under Geometrical Defects. These defects contribute to the variation of the part geometry from the nominal geometry (dimensions and form) and may also manifest as surface roughness and deviations on the plane surface. These can be detrimental to the structural integrity and may render the part unacceptable.

### 3.1.1 Non-uniform layer thickness:
**Case 1:** The deposited layer thickness is not constant and becomes uneven with the build height. This leads to a shorter or taller part.

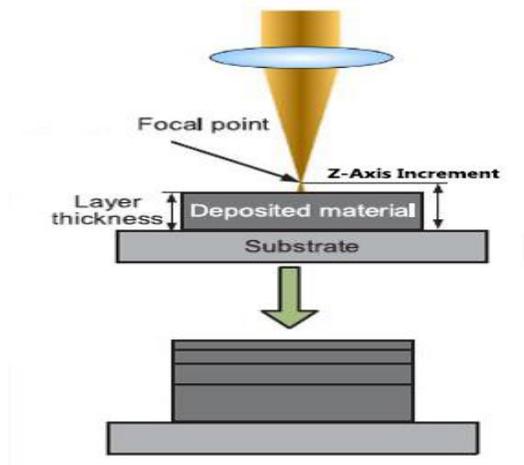

*Figure 4: Z-axis increment is greater than deposition thickness. Image is reproduced from [19].*

| CAUSE | MANIFESTATION | SOLUTION |
|---|---|---|
| Deposition thickness is smaller than the set z-axis increments due to non-optimal powder flow rate, too high scan speed, and effective focal length delivering too little powder. | Once the first layer has been deposited, this configuration results in a focal point that is above the top surface plane, thereby generating continually thinning layers [19, 20]. | Deposition thickness should be increased to bring it in line with z-axis increment by increasing the powder feed[19, 21]. Either the flow rate can be incresed, scan speed can be reduced or the program can be altered accordingly to set the z axis increment. |

**Case 2:** The consecutive layer thickness starts to increase with build height due to focal point anomalies.

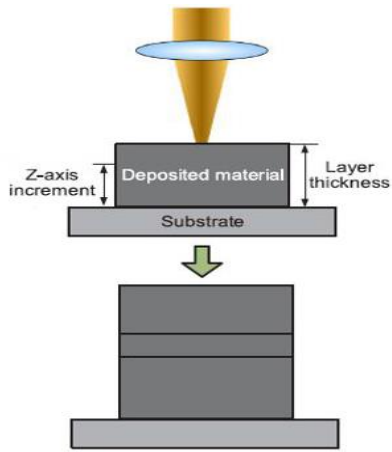

*Figure 5: Z-axis increment less than deposition thickness. Image is reproduced from [19].*

| CAUSE | MANIFESTATION | SOLUTION |
| --- | --- | --- |
| Deposition thickness is greater than the set z-axis increments due to non-optimal powder flow rate, too high scan speed, and effective focal length delivering too much powder. | It is manifested as generating continually thicker layers with build height. The z-axis increment being lesser then the resulting layer thickness results in a focal point that is below the top surface plane which lowers the incident laser energy density [19, 20]. | Deposition thickness should be reduced to bring it in line with z-axis increment by optimizing process parameter combinations (powder flow rate, scan speed, etc.). Either the flow rate can be incresed, scan speed can be reduced or the program can be altered accordingly to set the z axis increment. |

### 3.1.2 Warping on the top surfaces of the build:

The top surface of the build is not flat and exhibits intermittent peaks and valleys. The part appears to be deformed, along with changes in the aspect ratio. Problems arise from the inability to melt enough powder with a high scan speed. The scan speed, the powder feed rate, and the laser power all can contribute individually and in concert. Constant laser exposure reduces structural integrity due to the deepening of the melt pool as parts are built which leads to warping. Unmelted powder known as spatter is prevalent on the surface of the material. Materials that were made with higher laser power with the same scan speed lead may lead to a longer melt pool that collapses the structure. This leads to warping of the structure due to high energy output. The scan speed should be adjusted to properly melt all the powder. Less powder delivery could lead to an even melting and layered structure. Supports can be used to mitigate collapse by acting as a thermal sink.

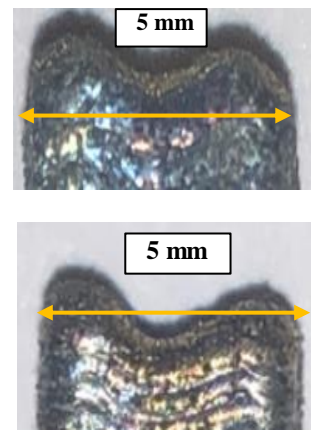

*Figure 6: Side view of a warped part made by LENS as reproduced from [17].*

| CAUSE | MANIFESTATION | SOLUTION |
|---|---|---|
| 1. Ineffective laser power<br>2. The scan speed and the powder feed rate<br>3. Deepening of melt pool | 1. Unmelted powder known as spatter is prevalent on the surface of the material.<br>2. A deepening melt pool leads to warping of the structure [1, 22].<br>3. Under excessive thermal stress, the top layers of the part may be deformed [23]. | 1. Adjust scan speed.<br>2. Support structure that acts as thermal sinks. |

### 3.1.3 Edge Defect:

The edge is not in level with the build plane and varies greatly from the as-modeled part. These might manifest as material overlay or underlay on opposite sides. A peak can develop at the start of the deposit with a valley forming at the end. As a result, poor part quality is observed specifically low density or part manufacturing failure. Optimized laser power, scan speed and powder feed rate help deposit the right amount of powder. Offset patterns from outside to inside and fractal patterns create lower thermal gradients, reduced substrate distortion, and improve component quality.

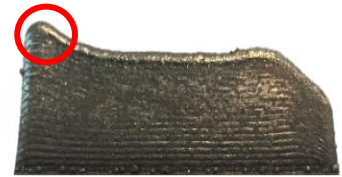

*Figure 7: Edge defect protruding as ridge on sample. Reproduced from [24].*

| CAUSE | MANIFESTATION | SOLUTION |
|---|---|---|
| 1. High laser power.<br>2. High scan speed.<br>3. Excess powder feed rate.<br>4. Scan patterns that induce high thermal gradients | 1. Peak and valley along one track [25].<br>2. Poor part quality [26]. | 1. Optimized process parameters [27].<br>2. Offset patterns from outside to inside and fractal patterns [24, 28]. |

### 3.1.4 Satellite Defect:

A surface satellite appears as a balled/rolled protrusion. This generally occurs in a region of overlap between the start point of the nozzle and the end of its path. These protrusions can be generally milled off. They are a combined effect of dragging and spatter. High scanning speed can deliver powder to unintended locations. When excess powder accumulates and assimilates to the surface due to laser, satellite defects occur.

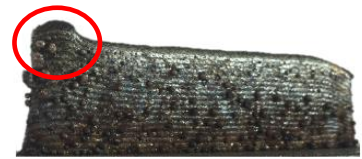

*Figure 8: Satellite defect protruding from the surface. Reproduced from [24].*

| CAUSE | MANIFESTATION | SOLUTION |
|---|---|---|
| 1. Non-optimal powder feed rate delivers too much powder contributing to spatter [29]. <br> 2. High scanning speed and accumulation of excess powder. | 1. Small round satellites appear as elevated regions of topography [24]. | 1. Consistent powder delivery can mitigate the incidence of uneven layers [30, 31]. <br> 2. Optimized scanning speed can ensure proper power delivery. |

### 3.1.5 Material Dilution leading to incorrect geometry of sample:

A LENS deposit may not have the correct geometry if a large degree of mixing occurs between the base substrate material and the deposited powder [42]. As the melt pool grows larger, it takes longer to solidify leading to a mixing between substrate and deposit known as dilution. The rate of dilution correlates to the size of the melt pool. Higher dilution of constituents in the melt pool during the fabrication process can lead to shorter and wider samples. To avoid dilution, an operator should utilize optimized process parameters through program control for each type of deposit rather than simply using constant process parameters.

| CAUSE | MANIFESTATION | SOLUTION |
|---|---|---|
| 1. Higher dilution of constituents in the melt pool [29, 32]. <br> 2. Melt pool size [33]. | 1. Dimensions larger than intended [34-36]. | 1. Utilizing program control methods [37-39]. <br> 2. Lower dilution [40, 41]. |

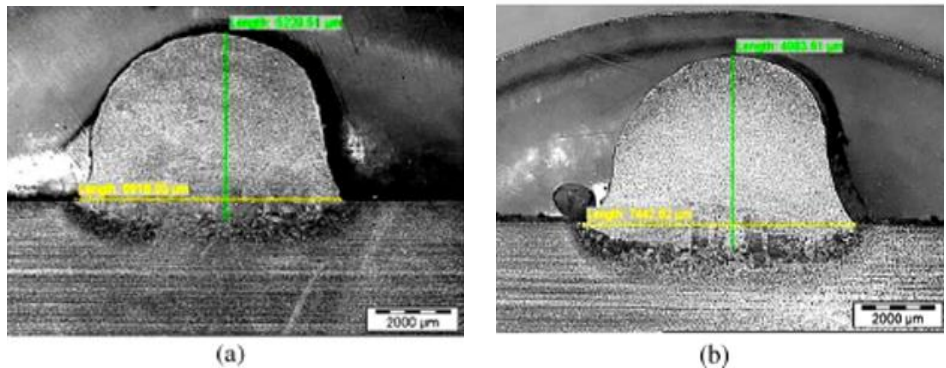

*Figure 9: Mixing between powder and melt pool creating longer melt pools that create incorrect geometry of sample [42].*

### 3.1.6 Material Shrinkage:

There is a negative offset which is less material deposition than the intended geometry. The part produced is thus smaller in size/dimension than desired. These can be rectified by depositing a compensatory layer to repair the part. Material shrinkage of a single layer during DED creates slanted edges or extremities around the layer itself while causing previously deposited layers to shrink via dragging. The Staircase effect results in poor surface quality and requires post-manufacture processing to form the desired shape and increases overall process time. Controlling layer thickness or volumetric difference between layers aids to reduce staircase effects.

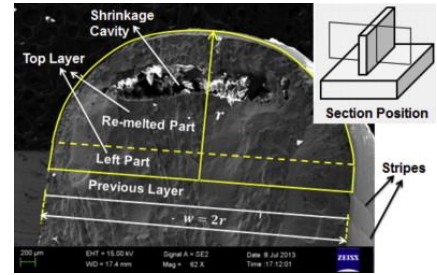

*Figure 10: Less deposition than the nominal geometry*

| CAUSE | MANIFESTATION | SOLUTION |
|---|---|---|
| Slanted edges or extremities around the layer causing previously deposited layers to shrink via dragging [43]. | 1. Staircase Effect 2. Poor surface quality that requires post-processing [29, 44, 45]. | 1. Controlling layer thickness or volume difference between layers 2. Multi-axis processing using AM can help mitigate the staircase effect [1, 46-48]. |

### 3.1.7 Warping at the bottom/separation from substrate:

The part deforms and buckles at the bottom and eventually detaches from the substrate which might render the part useless and cannot be salvaged by post-processing. This generally happens from excess heat input and absence of thermal channels to transfer heat.

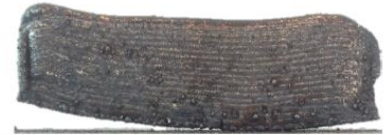

*Figure 11: Warping defect leading to detachment from substrate as reproduced from [24].*

| CAUSE | MANIFESTATION | SOLUTION |
|---|---|---|
| 1.Excessive heat input [24]. | 1.The sample detaches from substrate and results in a lack of fusion for the initial build | 1. Preheating of Substrate. 2. Optimization of dwell time [24]. |

### 3.2 Morphological Defects:

This type of defect, classified as either surface or bulk morphological defects, deals with the fabricated surface in terms of lay and feed patterns, texture, pores, cavities, and inclusions in the bulk.

### 3.2.1 Surface Morphology:

The morphological defects that appear on the surface that can be examined by microscopy.

### 3.2.1.1 Waviness:

The surface deviates from planarity and has wavy features on the surface. This affects the surface finish and increases the roughness values. [Inconsistent powder delivery] Powder spatter occurs in a wide array of directions which leads to uneven construction. Certain scanning patterns induces warping due to thermal gradients.

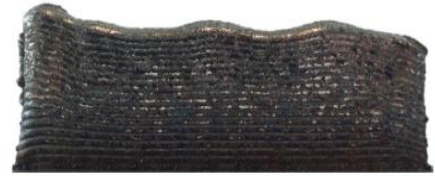

*Figure 32: Defect on the part called waviness. Reproduced from [24].*

| CAUSE | MANIFESTATION | SOLUTION |
|---|---|---|
| 1. Inconsistent powder delivery<br>2. Warp inducing heat due to thermal gradients [35]. | The layers are wavy and intermittent peaks and valleys can be seen [26]. | 1. Consistent powder delivery ensures proper layer evenness.<br>2. Parts can benefit from a zigzag scanning pattern with 90° change of orientation between layers [24]. |

### 3.2.1.2 High Surface Roughness:

Large variations in surface roughness can occur due to combined effects of several morphological defects. Rough surface features created by DED may act as stress concentrators and serve as origin points for potential failure in AM engineering components. This defect would be particularly amplified in thin members, where the effective surface layer is a larger proportion of the entire volume of the component. Detrimental features are mitigated through surface finishing. The surface has deviations from an ideal plane surface due to the combined effects of geometrical and manufacturing defects. Rework of the part through conventional machining techniques is required to achieve the desired finish. The surface shows uneven layers due to presence of fused unmelted powder stuck to the surface and heterogenous weld tracks. Precise control of process parameters can deliver uniform deposition and cooling rates.

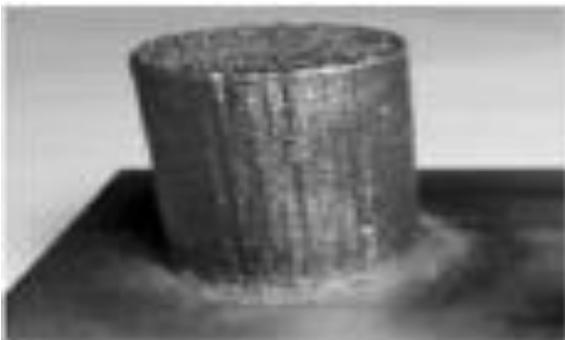

*Figure 13: Impaired surface finish of part developed by DED [49].*

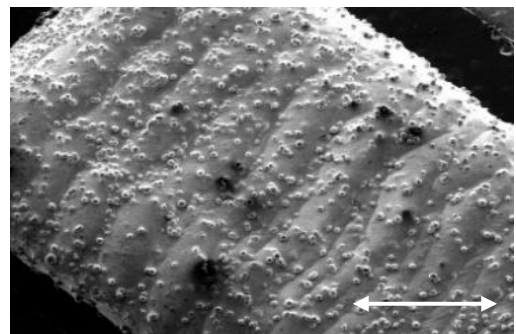

*Figure 44: SEM image of unmelted fused powder and uneven layers on the part side surface [17].*

| CAUSE | MANIFESTATION | SOLUTION |
|---|---|---|
| 1. Variation in powder size<br>2. Differential cooling<br>3. Unmelted fused powder on the surface<br>4. Surface deformation due to high heat input and scan pattern<br>5. Heterogenous weld track [24, 39, 49]. | 1. Rough surface features created during DED [24].<br>2. Thin member susceptibility<br>3. Undulations on the surface.<br>3. Waviness<br>4. Absence of Planar Surface<br>5. Unmelted fused particles [18] | 1. Fine surface finishing.<br>2. Control of process parameters for uniform deposition and cooling rate.<br>3. Post-print machining can bring part closer to net shape design. |

### 3.2.1.3 Blobs and Zits:

Blobs and zits may appear at the start or end point of the print when the laser is switched on or off thus creating an extra deposition. Overlap of the start and end point can cause this issue as the process cannot be seamless. The appearance of these surface blemishes can be minimized. These may appear as extra material generally at the starting and the ending point of a deposit where the melt pool overlaps.

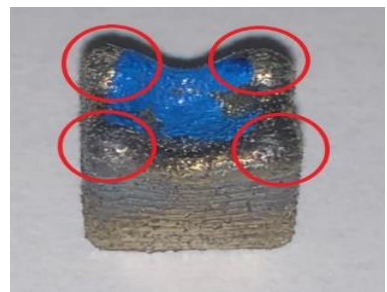

*Figure 55: Development of blobs and zits on the part [17].*

| CAUSE | MANIFESTATION | SOLUTION |
|---|---|---|
| 1. Unnecessary retractions and free run leads to accumulation of extra powder.<br>2. Inappropriate dwell time. | Extra material at the beginning and end of a deposit. | 1. Avoiding unnecessary retraction.<br>2. Optimized dwell time.<br>3. Proper runoff lengths.<br>4. Priming of the nozzle. |

### 3.2.2 Bulk Morphological Defects:

Defects that appear in the sub-surface are difficult to quantify unless metallography or sophisticated processes such as density measurements are performed.

### 3.2.2.1 Porosity:

Voids reduce material density and are present on surfaces and within the bulk. Several factors may favor the formation of pores within a part including gas entrapment, porosity in the powder, and lack of fusion [50]. Voids can lead to small cracks. This lowers the heat conduction ability of the material.

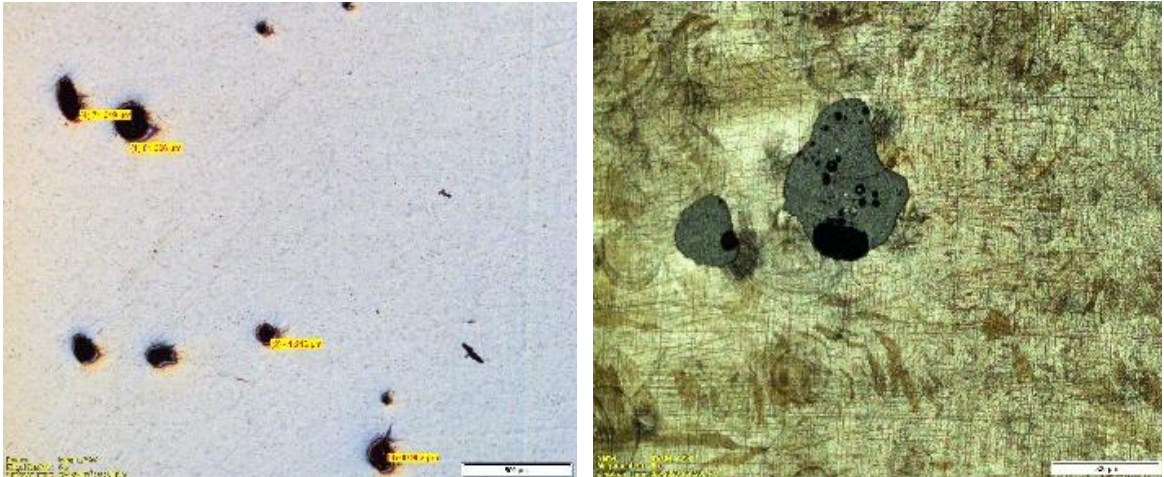
*Figure 66: Pores on parts developed by DED.*

| CAUSE | MANIFESTATION | SOLUTION |
|---|---|---|
| 1. Pores form due to gas trapping in the upper portions of the solidifying melt. 2. Pores may arise due to pores in the powder [51]. 3. It can also result from lack of fusion. | 1. Cracks form from some of these defects 2. Voids form during the initial build of the material. 3. Small cracks that lower heat conduction [37]. 4. Shorter fatigue life due to porosity and lower ductility [39]. | 1. Regulate flow rate of the fill gas and the pathway of the flow. 2. PREP (plasma rotate electrode process) powder yields less porosity than GA (gas atomized) powder [19, 29, 33]. 3. Using higher laser power and powder with fewer satellites will also help reduce the porosity level. |

### 3.2.2.2: Delamination/lack of adhesion between layers:

The adjacent layers delaminate upon application of load due to lack of adhesion between them. Cracking occurs due to thermal cycling during part building. Inadequate laser power can cause a lack of adhesion between layers. Residual stresses that are within the material can cause fracture. Fracture features show peeled-off regions between layers (in this case, coating layer). Reducing the formation of intermetallic that decrease structural homogeneity can help reduce the instance of delamination. Strategies from Section #3.2.2.1 can be used.

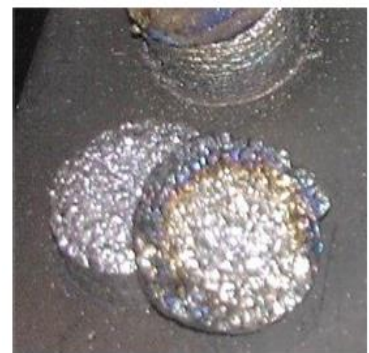
*Figure 77: Delamination of the part [52].*

| CAUSE | MANIFESTATION | SOLUTION |
|---|---|---|
| 1. Inadequate laser power<br>2. Residual stresses<br>3. Thermal cycling during part building [1, 53]. | 1. Part failure due to extensive cracking [53].<br>2. Delamination of structure due to excessive thermal stresses<br>3. Peeled-off and fractured regions between layers | 1. Control cooling rate to reduce the effects of thermal cycling and expansion differences of materials [54].<br>2. Adequate heat transfer channel.<br>3. Reduce the formation of intermetallics. |

### 3.2.2.3 Cavities:

Cavities or craters can form due to gas entrapment, inclusion of foreign particles, and lack of powder fusion. These defects act as stress concentrators and reduce the lifespan of the part. Cracks can originate from these cavities and lead to failure. Excessive amount of powder reduces the temperature of the melt pool and leaves some unmelted powder. An increase in laser power corresponding to the powder feed rate can ensure fusion. Strategies from Sections #3.2.2.1 and #3.2.2.2 can be used.

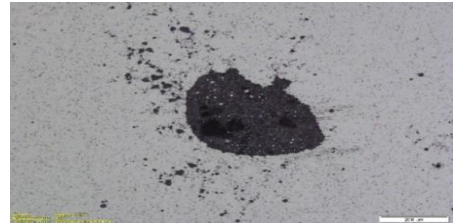

*Figure 88: Micrograph of the part produced using DED containing cavities.*

| CAUSE | MANIFESTATION | SOLUTION |
|---|---|---|
| 1. Gas entrapment in the melt pool during solidification can lead to cavities [51].<br>2. Lack of powder fusion creates a void.<br>3. Excessive amount of powder | Appears as voids from micrograph examination or superficial examination depending upon size. | 1. Increase laser power in line with powder feed rate.<br>2. Reduce the powder feed rate. |

### 3.2.2.4 Inclusions in the bulk:

Foreign particles implant themselves in the bulk of the part and degrade the material property due to lack of fusion. These foreign particles might originate from the powder or the DED machine. Unfused inclusions may concentrate stress and may act as originators of fracture.

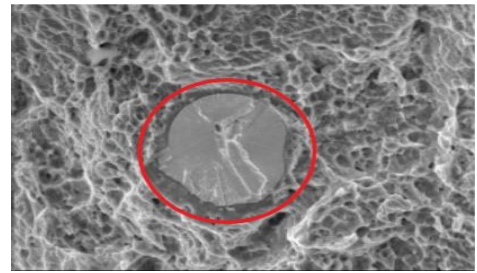

*Figure 99: SEM image showing impurity in the bulk [55].*

| CAUSE | MANIFESTATION | SOLUTION |
|---|---|---|
| 1. Cross-contamination from machine parts [56]. | 1. Stress begins to concentrate in the vicinity of these inclusions leading to fracture [56]. | 1. Cleaning the machine before the deposition. |
| 2. Inclusions in the powder. | | 2. Using contaminant free powder. |
| | | 3. Pre-Processing the powder to separate foreign particles. |

### 3.2.2.5 Non-uniform grain texture influenced by laser power:

The grain texture and orientation are affected due to differential cooling. This leads to inhomogeneity and anisotropy of properties. Parimi et al discovered that the columnar grain texture is significantly influenced by the laser power applied. For example, a microstructure may change from mixture of fine and coarse grains with weak texture to a fully columnar microstructure that is strongly textured. An adequate support structure can ensure uniform cooling. Optimized laser power reduces overheating and effects of temperature gradients.

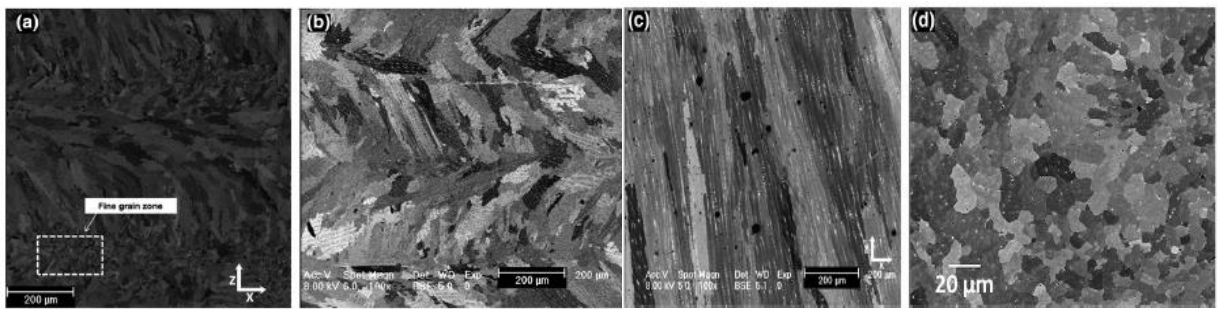

*Figure 20: BSE SEM micrographs showing multiple directions of grain structure influenced by heat gradient [57].*

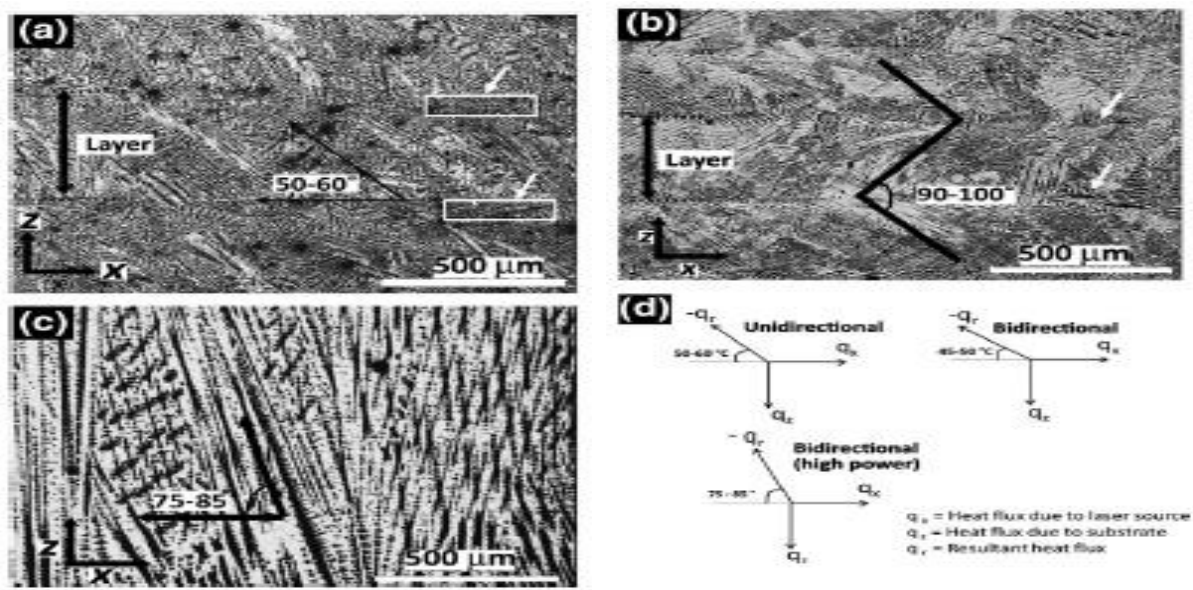

*Figure 101: Optical micrographs showing the dendrites orientation and layer demarcation in (a) unidirectional deposition, and (b) bidirectional deposition, using similar process parameters, (c) high power condition, and (d) schematic illustrations of the heat flux directions (not to scale). Reproduced from [57].*

### 3.2.2.6: Material susceptible to heat affected zone (HAZ):

Due to variation in scanning strategies and re-melting, intermittent HAZs form and govern the grain growth. Dilution from mixing powder and melt pool due to various HAZs creates composition variability over the part. For a bulk deposit (greater than 2 mm wide) primary solidification cell growth is influenced by the presence of interpass boundaries. High aspect ratio solidification cells appear to terminate at the interpass boundaries. Material is susceptible to a systematic interpass HAZ, which in most cases contains a mixture of recrystallized coarse grains and/or poorly defined soft material zone due to a reheating and re-melting from the successive layer deposition. Interpass HAZ type defects appear in bulk DED deposits, particularly in the areas surrounding un-melted/partially melted particles. To remedy this, controlling grain growth and increasing lamellar width or grain width during heat treatment can tailor microstructure formation and reduce material property differences.

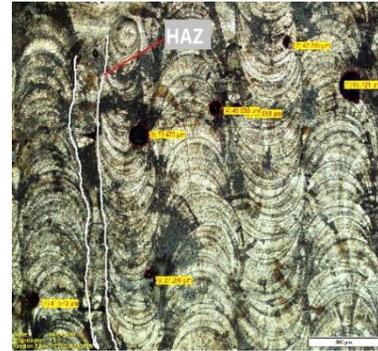

*Figure 112: Micrograph showing weld tracks and HAZ.*

| CAUSE | MANIFESTATION | SOLUTION |
| --- | --- | --- |
| Reheating and re-melting from successive layer deposition leads to a HAZ that creates composition variability. | 1. Material exhibits a mixture of recrystallized coarse grains and/or poorly defined mushy zone. 2. HAZ type defects in bulk deposits, in areas near un-melted/partially melted particles [57]. | Control of the grain growth and increasing lamellar width or grain width during heat treatment [58, 59]. |

### 3.2.2.7 Lack of powder fusion:

Some of the unfused powder might remain implanted in the bulk due to suboptimal laser power or excess powder feed into the melt pool. These unfused trenches act as stress concentrators and might favor crack initiation and propagation. Excessive powder flow rate can cause too much powder to shoot into the melt pool and simply stick to adjacent areas or as the part is being built will remain stuck to the final structure. The problem manifests as unmelted powder on the surface and within the sample in the form of pores or lack of fusion trenches. Use of higher power and lower scan speed can be instrumental in producing better prints. In the case of a study performed by authors, shown in *Figure 25* using 250W-300W and a slower scan speed of 3-5 mm/s should be able to mitigate these effects. Further strategies can be employed from Sections #3.2.2.1-3.2.2.3.

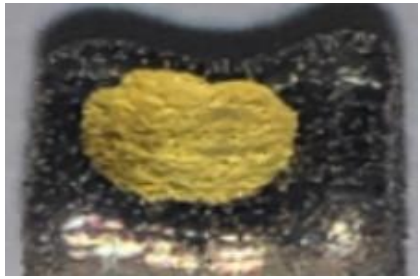 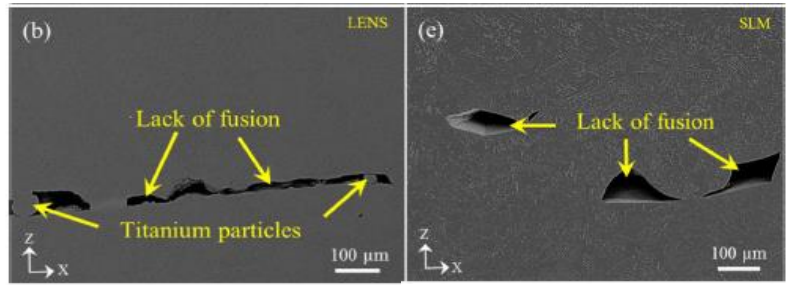

*Figure 12: Lack of powder fusion on surface of sample[17].*

*Figure 13: Lack of fusion defects in CP-Ti: porosity (left) and trenches (right) [68].*

| CAUSE | MANIFESTATION | SOLUTION |
|---|---|---|
| Excessive powder flow rate can cause too much powder to shoot into the melt pool and simply stick to adjacent areas [60]. Lack of energy input may lead to insufficient melting. | Unmelted powder on the surface and within of the sample in the form of pores and lack of fusion trenches.[63, 64]. | Use of higher power and lower scan speed for better prints. [61]. |

**3.2.2.8 Incomplete melting in transition layers for a compositional gradient:**
Due to differences in powder composition at the interfacial regions and stochastic nature of dissimilar metal fusion there might be a lack of adhesion between layers which can favor delamination. Insufficient power or lack of fusion between dissimilar materials can be a cause. The surface contains un-melted or partially melted particles, which are fused to the deposit surface and tend to gather along the molten metal flow trails and/or interpass boundaries. Unmelted solid inclusion can interrupt metal flow. Interrupted metal flow appears to change the heat transfer and flow dynamics of the process, resulting in the formation of recrystallized grains within HAZs. Fused-on un-melted powder particles not only modify the ultimate surface roughness, but also become a source of interpass inclusions if they remain unmelted and ultimately the root cause of adverse interpass porosity. Preheating powders before delivering mixtures can mitigate these effects. Increasing the liquid metal pool temperature in the first layer of transition region can allow for smoother layer deposition. Finer powder size can help improve melting. Exothermic mixing between two major constituents can gradually increase the melt pool temperature. Strategies from Sections #3.2.2.1-3.2.2.3 and #3.2.2.7 can be used.

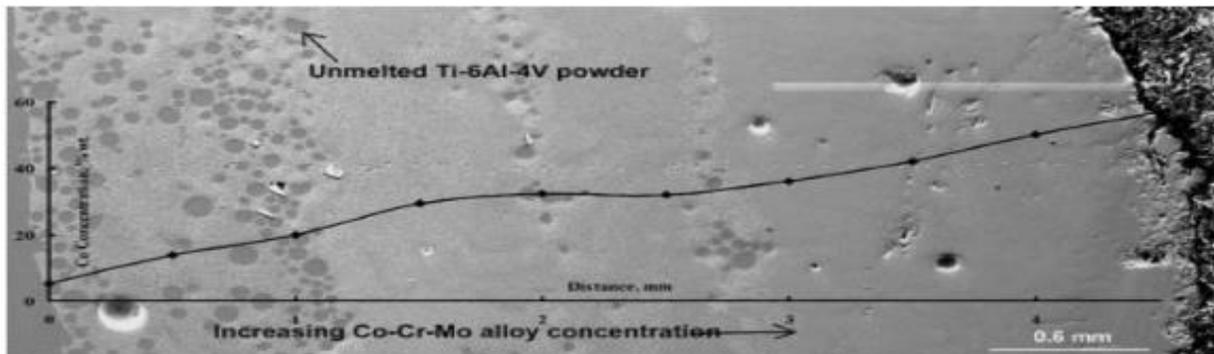

*Figure 25: Typical microstructure and Co distribution of 86% Co-Cr-Mo graded coatings on porous Ti-6Al-4V alloy using LENS. Adapted from [15].*

| CAUSE | MANIFESTATION | SOLUTION |
|---|---|---|
| 1. Insufficient power or lack of fusion [60]. 2. Structure shows compositional inhomogeneity which leads to incomplete melting [61]. | 1. Formation of recrystallized grains within HAZs 2. Unmelted solid inclusion can interrupt metal flow. 3. Increased surface roughness due to particles that exist within the zone between two materials [49]. | 1. Preheating powders before delivery [23]. 2. Finer powder size for facile melting 3. Increasing the liquid metal pool temperature in the first layer of transition region. 4. Exothermic mixing [61]. |

### 3.3 Microstructural Defects:

Probing into the microstructure reveals essential information about the dynamics of the deposition process and material integrity and behavior. The spatial arrangement of grains and segregation of elements can be analyzed. This class of defects furthers our insight on the causal relationship to the process variables and allows us to think of novel solution to mitigate them.

### 3.3.1 Feature-Related Defects:

These defects appear as undesired features in the microstructure which affects the characteristics and mechanical properties.

### 3.3.1.1 Discoloration:

Overheating and differential cooling of the bulk is manifested as a spectrum of color on the surface. Abnormal color may precede the development of unwanted oxides [18]. High thermal gradients in the LENS process may produce different microstructural regions: columnar at the bottom and equiaxed at the top. Problem may manifest as a discoloration in the part where the reheating occurs repeatedly while a large amount of powder is being shot as shown in the samples shown above. Mitigating discoloration and its detrimental effects can be achieved through careful attention to the process parameters.

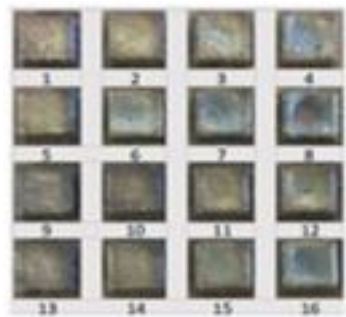

Figure 146: The dimension and shaping of the deposited blocks [62].

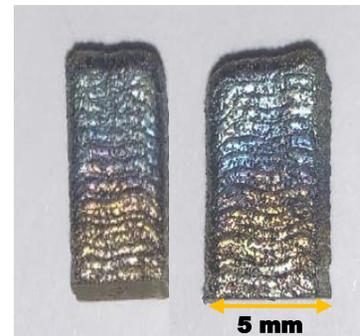

Figure 157: Discoloration of the sample produced by DED [17].

| CAUSE | MANIFESTATION | SOLUTION |
|---|---|---|
| 1. Overheating from temperature gradients. 2. Intermetallic formation | 1. Discoloration [63]. 2. Brittle failure [63, 64]. | Careful attention to the process parameters [62, 65]. |

### 3.3.1.2 Microstructural non-homogeneity:

The heterogeneous microstructures of metal DED vary in grain type and size based on cooling rate and thermal history. This heterogeneity leads to anisotropic mechanical properties like inconsistent hardness values. Cooling rate of the melt pool dictates the microstructure formed in a deposited layer. Higher cooling rates result in finer microstructures. Some indications of microstructure inconsistency include the appearance of equiaxed, columnar dendritic growth and epitaxial structure. Even cooling rate can be achieved by imbibing thermal sink in form of support structures. Optimized laser dwell time can reduce instances of directional grain growth. Strategies from Section #3.2.2.5 can be used. Due to temperature gradients that exist within LENS during its run time, parts may develop anisotropy in the microstructure. The power input and consistent heat cycling cause variable microstructures to form. Microstructure changes with the rate at which the heat sinks into the substrate at the bottom of the build. Microstructure shows variation based on location. For example, some builds exhibit equiaxed grains at the top of the build, a region exposed longer to cyclic reheating. The bottom of the build is made up of columnar grains. Transitioning from columnar to equiaxed grain structures can be attributed to increase in solidification rates. These different regions assume different mechanical behavior and adds to the stochastic character of the build. A controllable cooling mechanism that reduces the effects of heat cycling should help create a consistent columnar structure.

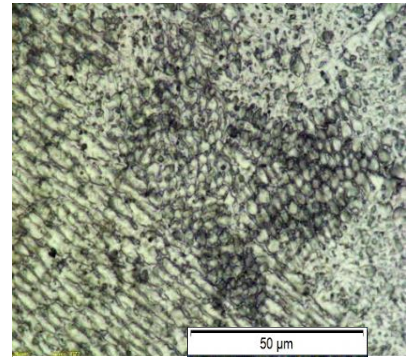

*Figure 28: Micrograph of SS316LN developed by DED showing non-homogeneous microstructure.*

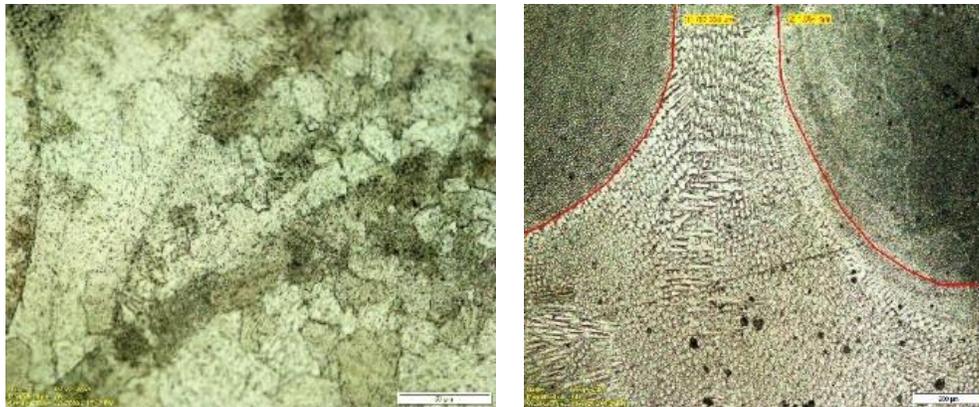

*Figure 29: Micrograph showing non-homogenized grain growth (left), presence of columnar and equiaxed grains (right).*

| CAUSE | MANIFESTATION | SOLUTION |
|---|---|---|
| 1. Non-homogeneous cooling rate of the melt pool [66]. <br> 2. The power input and consistent heat cycling <br> 3. Heat sinking into the substrate at the bottom of the build [67]. <br> 4. Solidification rates | 1. Appearance of equiaxed, columnar dendritic growth and epitaxial structure [66]. <br> 2. Microstructure shows variation based on location. <br> 3. Transition from one structure to another [68]. <br> 4. Variable mechanical behavior [69]. | 1. Even cooling rate through imbibing sink [66]. <br> 2. Optimized dwell time <br> 3. A controllable cooling mechanism <br> 4. A rotating arm that can move the piece and direct laser energy to avoid heat cycling. |

### 3.3.1.3 Inter- and intra-cracking in the structure:

Cracks in the part might be present because of lack of fusion or thermal stress. Pores in the part act as crack initiators and favor propagation. These factors reduce the fatigue life and might lead to part failure [66]. Interlayer cracks form due to solidification along grain/dendrite boundaries and stresses that develop when the material shrinks. Optimized process parameters yield crack-free deposits with minimal porosity. Strategies from Sections #3.2.2.1-3.2.2.2 can be used.

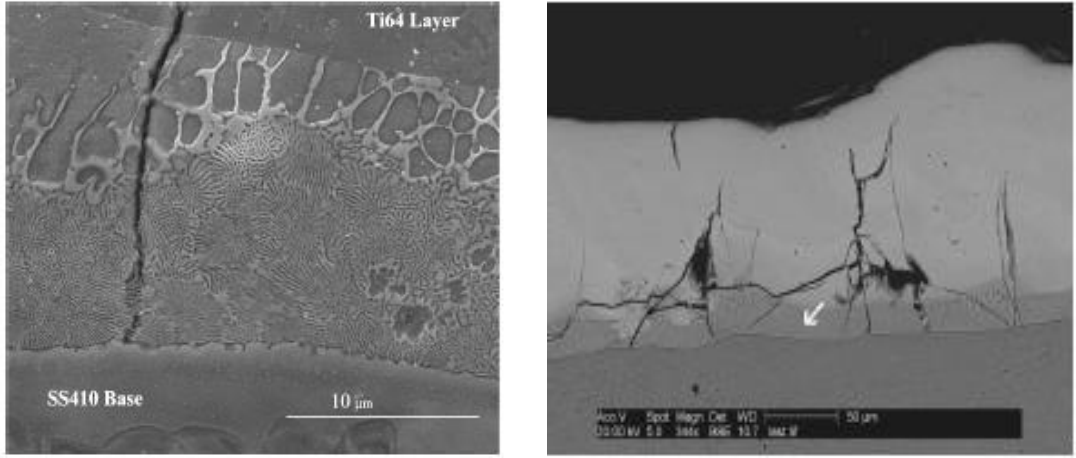

*Figure 16: Cracks from AM shown across layers (left) and within a layer (right).*

| CAUSE | MANIFESTATION | SOLUTION |
| --- | --- | --- |
| 1. Solidification of continuous films along grain/dendrite boundaries<br>2. Solid shrinkage stresses [29]. | Intra-layer cracks occur through individual layers [70]. | 1. Optimized process parameters [66].<br>2. Increse in scan speed.<br>3. Increase in laser power. |

### 3.3.1.5 Tensile Strength Anisotropy:

The tensile strength varies with the direction of build and scanning strategy. Different values are measured for distinct length and direction of columnar grains. Basketweave microstructures have different properties than those of dendritic structures. Anisotropy may occur due to the existence of non-aligned columnar grains. α layers at columnar grain boundaries are subjected to accelerated damage when tension is applied perpendicularly to the columnar grains. The α morphology and texture also influence the tensile behavior. The basketweave microstructure shows higher strength and ductility due to the randomly oriented α laths, which effectively decrease the dislocation slip length and reduce local stress concentration. Localized heat transfer, especially in the HAZ, can result in layer re-melting and numerous reheating cycles. This can change the microstructure by tempering and aging the part. Strategies from Reference #3.2.2.5 and 3.3.1.4 can be used.

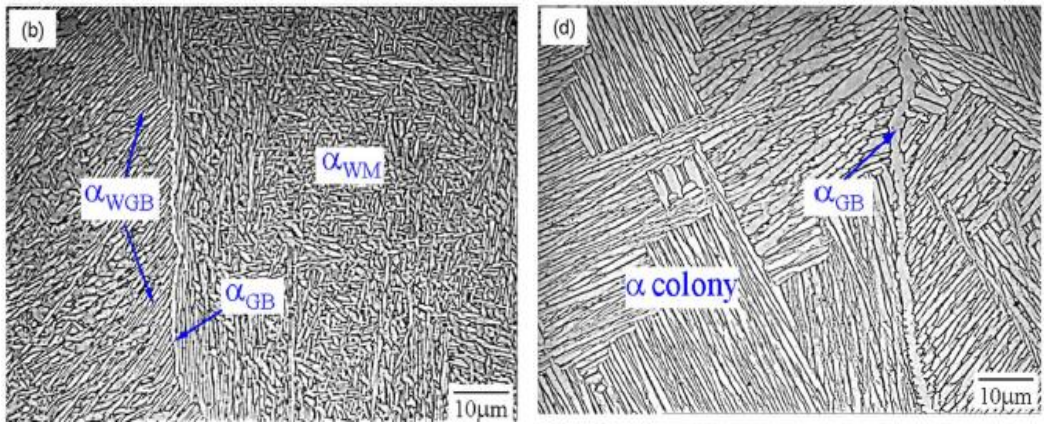
*Figure 17: Basketweave microstructure as shown in [71].*

| CAUSE | MANIFESTATION | SOLUTION |
|---|---|---|
| 1. Existence of non-aligned columnar grains [66].<br>2. Microstructures have different properties.<br>3. Morphology and texture | 1. α layers at columnar grain boundaries are undergo accelerated damage when tension is applied.<br>2. Basketweave microstructure shows higher strength and ductility<br>3. Decrease in dislocation slip length and local stress concentration [15]. | 1. Localized heat transfer, especially in the HAZ, can result in layer re-melting and numerous reheating cycles.<br>2. Tempering and aging the part. |

### 3.3.2 Compositional Defects:
These defects are due to heterogeneous elemental overlay and dilution which leads to the formation of unwanted intermetallics and phases.

### 3.3.2.1 Development of unwanted intermetallic compounds:
Unwanted intermetallics are sometimes formed during the introduction of powder into the melt pool and dilution which degrades the material quality in terms of stability and mechanical integrity. This defect manifests as intermetallics that are detrimental to the final material due to the heat input and constituents. The formation is governed by the thermal history of the part. When intermetallic compounds are formed, they can change material properties that are unfavorable to the product. The combination of scan strategy and process parameters must follow a certain compositional and temperature path based on phase diagrams to avoid the intermetallic phases from forming.

| CAUSE | MANIFESTATION | SOLUTION |
|---|---|---|
| Thermal history of the part due to heat input and constituents [64]. | Reduction in material properties [57]. | Combination of scan strategy and process parameters. |

### 3.3.2.2 Development of undesirable phases that lead to reduced material properties:

Development of phases, such as brittle Laves phase, can reduce the life cycle of the component. These detrimental phases decrease the tensile strength of the developed part. LENS Ti-6Al-4V has much lower ductility due to the presence of martensitic α' phase. Higher power LENS Ti-6Al-4V has slightly lower strength and higher ductility due to a lower fraction of martensite compared to the low power condition. Columnar β grains form as result of heat extraction from the substrate. Laves phase forms during DED manufacture of IN718 which consists of columnar grains within Nb-depleted austenitic dendrites with fine secondary dendrite arm spacing (SDAS). Segregation of Nb, Mo, and Ti elements results in the precipitation of brittle Laves phase at the inter-dendritic regions. For LENS, a simple post fabrication heat treatment that is suitable for repair can be used as long as the original microstructure is kept intact. Preheating the substrate allows for stress relief and decomposition of undesirable phases. For Laves, process optimization and post process heat treatments reduce amount of segregation and Laves phase. Xiao et al used a quasi-continuous-save process and achieved much faster cooling rates during fabrication, dramatically reducing the size of SDAS and the amount of Nb segregation. Solution and aging-based heat treatment can reduce the concentration of Laves phase and increase material tensile strength.

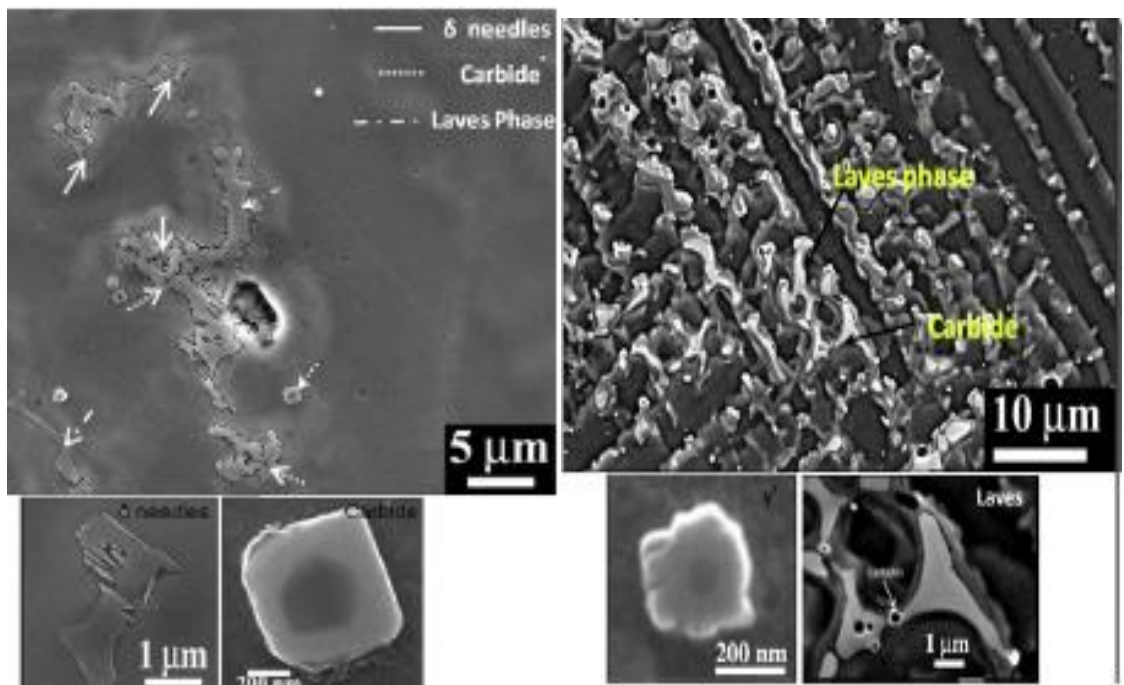

*Figure 18: BSE SEM micrographs showing the intermetallic precipitates in (a) B2 (lower inset shows the carbide and Laves phases at higher magnification) and (b) B3 (lower inset shows the δ needles at stacking faults and carbides at higher magnification) [68].*

| CAUSE | MANIFESTATION | SOLUTION |
|---|---|---|
| 1. Non-optimal power contributes to undesirable phases [53]. 2. Heat extraction from the substrate. 3. Segregation of elements | 1. Causes reduced material properties [18]. 2. Precipitation of brittle Laves phase at the inter-dendritic regions [35, 36, 72-75]. | 1. A simple post fabrication heat treatment [76]. 2. Preheating the substrate [77]. 3. For Laves, process optimization and post process heat treatments [18]. 4. Quasi-continuous-save process to achieve much faster cooling rates during fabrication [69]. 5. Solution and aging heat treatment to reduce concentration of Laves phase. |

### 3.3.2.3: Elemental segregation:

Due to different dilutions over the part, the elemental composition varies non-uniformly over the part which develops regions with varying properties. Interpass HAZ may contribute to dilution by increasing length of melt pool. Dilution of parts can cause isolation of elements. EDS analysis of these parts shows the redistribution of elements to different regions reflecting the variation.

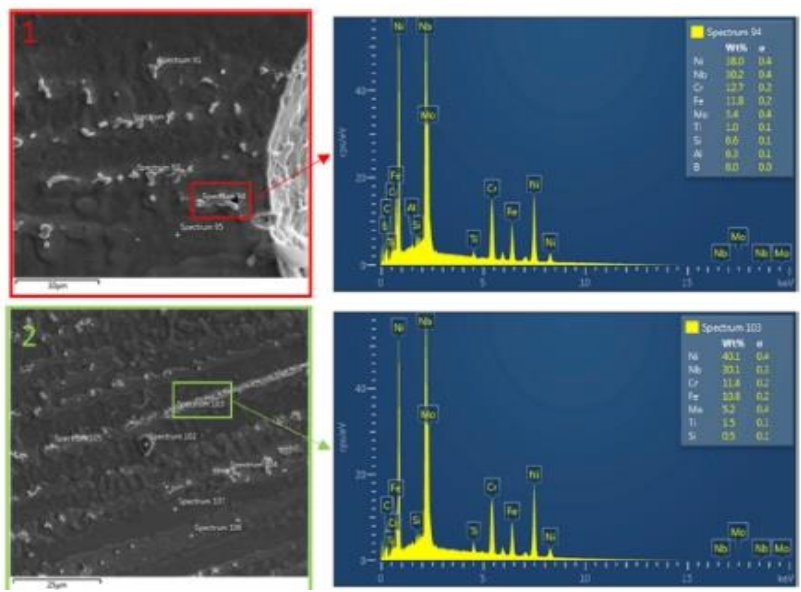

*Figure 19: EDS analysis showing inconsistent elemental concentration [69].*

Redistribution of elements and loss of surface properties may ensue. Post processing and heat treatment can allow proper mixing. Low heat input can also ensure integration.

| CAUSE | MANIFESTATION | SOLUTION |
|---|---|---|
| 1. Interpass HAZ 2. Formation of different phases 3. Dilution of parts | 1. Redistribution of elements. 2. A loss of surface properties | 1. Post processing and heat treatment 2. Low heat input |

# 4. PROCESS AND MACHINE TOOL FAULTS:

## 4.1 Anomalous machine or process settings:

Optimal process parameters are the recipe to successful prints. Classification of this type of defect helps further insight about how to avoid these anomalies and develop corresponding manufacture paths and algorithms. Info from the feedback and maintenance systems can provide correctional guidance. In many cases, constant monitoring from operators can compensate for faults in the system.

### 4.1.1 Process Settings:

#### 4.1.1.1 Powder Over/Under Feed:

Powder feed rate should be commensurate with the laser power and scanning speed. Excess powder delivery can lead to unfused material which is detrimental to the material integrity. Sub-optimal powder delivery leads to excessive heat-related issues such as causing geometrical defects as buckling, waviness, etc. Anomalous powder feed rate can be due to non-optimal feed rate settings or due to inadvertent faults in the supply line or nozzle like clogging of nozzle, design of the nozzle aperture, leak in the powder feed line or gas supply. Overheating of nozzle may lead to non-optimal powder delivery. Clogging of the nozzle may lead to powder accumulation and uneven distribution.

| CAUSE | MANIFESTATION | SOLUTION |
|---|---|---|
| 1. Overheating of nozzle | 1. Material Overlay | 1. Using optimal purging gas flow [78]. |
| 2. Clogging of the nozzle | 2. Material Underlay | 2. Avoiding nozzle overheating |
|  | 3. Material Shrinkage | 3. Using correct nozzle aperture, working distance, and laser focus alignment. |

#### 4.1.1.2 Anomalous Laser Power:

Anomalous laser power can lead to shortening the depth or widening the area of the melt pool which greatly affects the build quality. The anomalies in the laser power could result from a plethora of sources like sub-optimal process setting, laser attenuation from the artifacts of the build and precursor material or offset error which changes the laser focus [67]. If the z-axis increment is not commensurate to the deposition thickness, the geometry of the part can be affected. This issue exacerbates with increasing build height. Laser attenuation may cause optical and energy density issues. Laser focus offset can contribute to improper energy density. Less than optimal power leads to lack of fusion defects. Excess heat input leads to buckling and other geometrical defects.

| CAUSE | MANIFESTATION | SOLUTION |
|---|---|---|
| 1. Laser attenuation [37, 43, 79]. | 1. Lack of fusion defects [30, 81]. | 1. Deposition thickness should be brought in line with z-axis increment. |
| 2. Laser focus offset [80]. | 2. Buckling and other geometrical defects. | 2. Powder shot direction should be offset with laser shot direction. |

### 4.1.1.3 Gas flow rate:

Excess gas flow rate can lead to increased gas entrapment which leads to porosity. Sub-optimal flow rate can lead to inept shielding and hence oxidation of the melt pool and part. Gas flow dynamics can influence discoloration of the part as a visible derivative of oxides. Inconsistencies across the build plate correlating with the gas flow direction can play a role. Anomalies in gas flow can arise by virtue of incorrect setting, sub-optimal parameter, erroneous gauges, and faults in the supply line.

| CAUSE | MANIFESTATION | SOLUTION |
|---|---|---|
| 1. The amount of powder that is delivered into the melt pool varies according to the flow rate of the shielding gas [82]. 2. Erroneous gauge reading. | 1. Higher amount of porosity in the component may result from trapped gas. 2. Discoloration can result from gas flow dynamics. 3. High gas flow rate increases outreach of particles which affects deposition. 4. Varied gas dynamics at melt pool can cause structural integrity issues. | 1. Gas flow should be monitored and adjusted to avoid sub-optimal gas flow. 2. Gas outlet diameter should be increased to reduce porosity. |

### 4.1.1.4 Scanning strategy anomalies:

Scanning strategies affect the final properties and amount of residual stress within the component after the deposition [84, 85]. Dai and Shave reported that by using the offset-out strategy it would be possible to reduce the residual stress to one third of the one produced by the bi-directional scanning strategy [84]. On the other hand, Nickel et al. have found that by using the raster strategy, the simplest and most common strategy, and rotating layers of 90° for alternating layers it would be possible to build different components with less part deflections [86]. The selection of the deposition strategy is still a key challenge for complex geometries. Fractal and offset strategies could attract more attention owing to their features of geometry accuracy and less energy consumption [87]. Heat treatment can help create a desired microstructure. Post-processing of the component can reduce the instance of geometric accuracies.

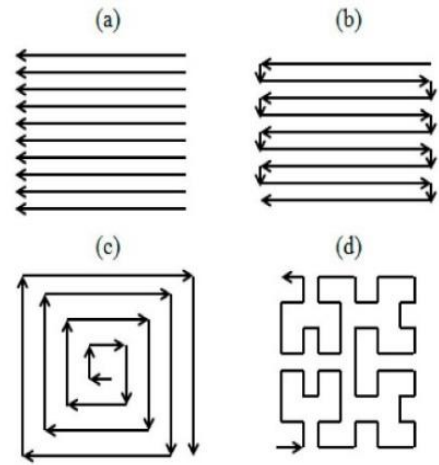

Figure 20: Different deposition patterns: (a) raster, (b) bi-directional, (c) offset-out, and (d) fractal [83].

| CAUSE | MANIFESTATION | SOLUTION |
|---|---|---|
| 1. Improper scanning strategy 2. Inattention to dynamics of scan path | 1. Residual stress within the component [88, 89]. 2. Geometrical inaccuracies. 3. Anisotropic properties [87]. | 1. Heat treatment 2. Post-processing |

### 4.1.1.5 Incorrect powder feeder RPM calibration:

Improper correlation and measurement of powder feeding flowrates alongside various powder feeding pressures and rotary speeds can lead to calibration problems. At high rotation speeds for the powder feeding disk, metal powder deposition occurs in the pipeline. Because of the pressure fluctuation change, the powder agglomeration, etc., the relationship between the actual powder feeding rate and the rotary speed is indeterminate. In the case of constant outlet pressure of throttling valve of gas bottle, with the increase of rotary speed of powder feeding disk, the pressure at the outlet of powder feeding pipe shows the upward tendency overall. To better realize the stable control of powder feeding flowrate, the outlet pressure of gas bottle should choose the middle grade. Then, the powder feeding flowrate could be adjusted by controlling the rotary speed of powder feeding disk. The powder flow rate in powder feeding tube can be confirmed through the testing calibration system. In the case of accumulation in the pipeline, the rotatory speed must be raised.

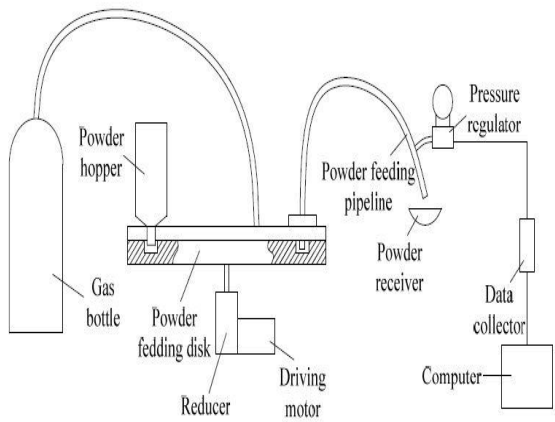

*Figure 21: Theoretical diagram of powder flowrate calibration system [90].*

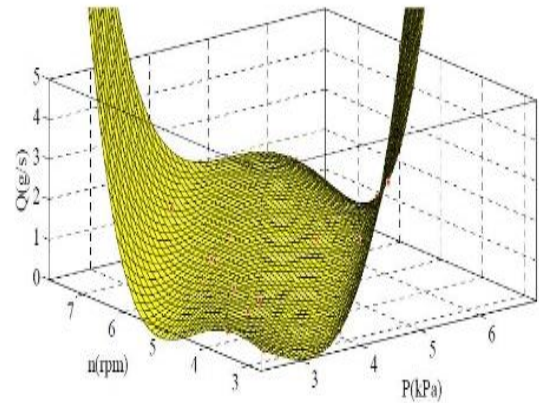

*Figure 22: Relationship between powder feeding flowrate and pressure, rotary speed through 2D fitting and testing data points [90].*

| CAUSE | MANIFESTATION | SOLUTION |
|---|---|---|
| 1. Improper correlation and measurement of powder feeding flowrates [90]. 2. Indeterminate relation between powder feeder disk and gas flow. | 1. Actual powder feeding rate and the rotary speed is indeterminate [90]. 2. The pressure at the outlet of powder feeding pipe shows the upward tendency overall. | 1. The outlet pressure of gas bottle should choose the middle grade. Then, adjusting powder feeding flowrate [90]. 2. Confirmation through the testing calibration system. 3. Raise in rotary speed instead of gas flow can be instrumental in providing better powder delivery for certain cases. |

### 4.1.1.6 Build direction anomalies:

The part can be built in different orientations. A certain orientation can be advantageous in lieu of support requirements, though mechanical properties vary along a direction. Variance in thermal history and cycling affect the microstructure and directional growth. Different cooling rates may cause microstructure orientation differences. The best resort is to build in the direction which asserts best mechanical property in the direction of loading. Post-processing methods like hot isotactic pressing and heat treatments can homogenize the microstructure to yield similar properties.

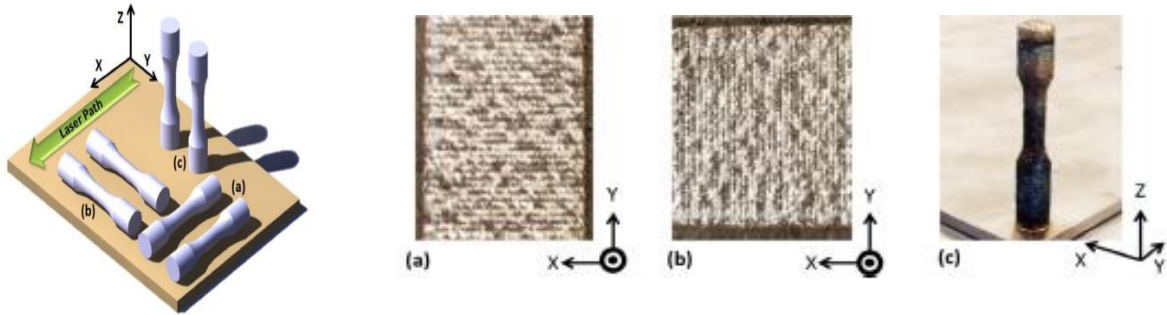

*Figure 23: Schematic view of DED Ti-6Al-4V specimens, (a) build direction of specimen in the X-direction (b) Y-direction, and (c) Z direction [87].*

| CAUSE | MANIFESTATION | SOLUTION |
|---|---|---|
| 1. Variance in thermal history and cycling [91]. 2. Different cooling rates. | 1. Anisotropic microstructure [26]. 2. Anisotropic mechanical behavior [68, 87]. | 1. Heat treatment helps create equiaxed microstructure [35, 87]. |

### 4.1.2 Latent State:

These conditional states cannot be changed and tracked in real time and has a bearing on the process dynamics to a great extent although these maybe be ascertained through a set of data. Most of them are not monitored and these include the properties of the stock material used in the process and machine specifications, among others. Certain other pre-defined factors that determine the initial conditions of the process environment and the real-time process state that need to be monitored; these include the chamber conditions, gas flow, temperature, etc.

### 4.1.2.1 Distance between the powder stream and laser power:

The relative direction of movement and placement of the powder nozzles and laser beam determines where the powder is delivered. If the powder stream is not in line with the laser stream it causes laser attenuation and erratic deposition. When the powder is delivered to point A is ahead of the laser spot O, the powder introduced to the melt pool is lower than when powder is introduced behind the laser [87]. The radial distance between powder stream and laser beam depends on the machine specification and the number of nozzles and nozzle orientation. Different scanning strategies have varied effects on the powder shot direction and the movement and spatial position of the melt pool. To have a constant mass flow rate, the powder feed rate and laser scan speed should be set according to the laser scan direction and the distance between the laser spot and nozzle.

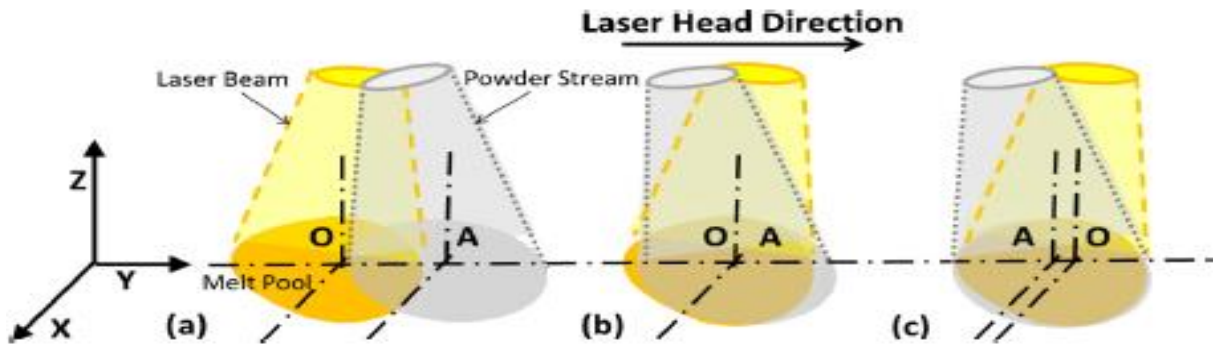

*Figure 24: Powder delivery position A (a) ahead of, (b) in-line with and (c) behind the laser spot center [92].*

| CAUSE | MANIFESTATION | SOLUTION |
|---|---|---|
| By changing the direction of laser scanning the point of powder delivery changes and hence the melt pool dynamics and can be ahead, in line or behind the laser spot [1, 93]. | This variation affects the geometry of the melt pool, boundary, and solidification and thus the height of deposition. | Powder feed rate and laser scan speed should be set according to the laser scan direction and the distance between the laser spot and nozzle [20, 83]. |

### 4.1.2.2 Overheating of Nozzle:

The nozzles are susceptible to overheating due to their proximity to the weld pool and sub-optimal cooling settings. The heat radiating from the melt pool can overheat the nozzle which alters the powder flow characteristics. Excess heat can lead to expansion of the nozzle and change the nozzle aperture. To avoid overheating of the nozzle, its temperature should be monitored through thermal sensors. A leak in the gas supply lines can also lead to inefficient cooling. Printing multiple objects at once allows for more cooling time for each object because the nozzle will be moving to different location to print different components which provides an opportunity to cool between laser pulses. Changing the distance between prints can affect the amount of cooling the nozzle undergoes. This is a simple and effective strategy to mitigate the overheating.

| CAUSE | MANIFESTATION | SOLUTION |
|---|---|---|
| 1. Insufficient heat dissipation can cause nozzle overheating.<br>2. Printing temperature is too high<br>3. Excess heat and laser power contribute to nozzle overheating due to proximity. | 1. Clogging of nozzle.<br>2. Erratic deposition.<br>3. Material overlay<br>4. Material underlay. | 1. Increase the cooling and purging gas flow.<br>2. Printing multiple objects at once allows for more cooling time |

### 4.1.2.3 Laser attenuation:

If the powder stream merges and intercepts the laser stream before their incidence on the focal plane, attenuation of the laser power occurs and alters the energy density. The spatial orientation of laser stream and powder stream is typically pre-defined by the machine manufacture, but attenuation can result from artifacts and precursor materials in the system. Erroneous offsets can cause merging of the powder and laser stream before their incidence on the focal plane.

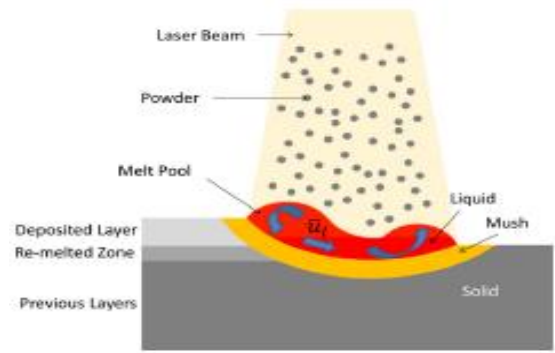

*Figure 25: Schematic of powder introduction to a melt pool [1].*

| CAUSE | MANIFESTATION | SOLUTION |
|---|---|---|
| 1. Scattering of laser ray by shot powder [1]. <br> 2. Misaligned nozzle. | 1. Lack of fusion <br> 2. Diffused laser spot which effects the energy density. | Directing the feed nozzle away from laser spot. |

### 4.1.2.4 Nozzle clogging:

The powder delivery nozzle can be clogged due to changes in effective nozzle aperture such as non-spherical powder particles and coalesced stock powder lobes which can induce blocks. These issues can alter the powder feed rate and the direction of the powder stream. Overheating of the nozzle and impurities may cause clogging which can block the powder supply line. The defect in the powder feed rate renders non-homogeneous deposition. This can lead to sub-optimally or over dispensed powder and formation of blobs. A heat sink for the nozzle can increase the heat release and reduce overheating of the nozzle. Further, a recirculating coolant system can help mitigate these issues.

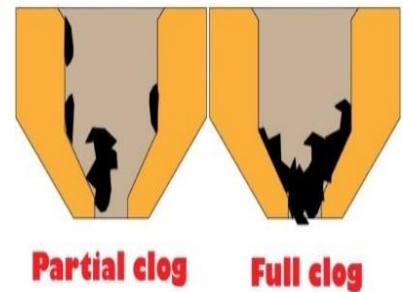

*Figure 26: Schematic image depicting partial or full clog.*

| CAUSE | MANIFESTATION | SOLUTION |
|---|---|---|
| 1. Overheating of nozzle. <br> 2. Non-Uniform particle size. <br> 3. Coalescing of powder. <br> 4. Foreign particle. <br> 5. Incorrect nozzle aperture. | 1. Inefficient powder delivery. <br> 2. Erratic deposition pattern <br> 3. Missed prints. | 1. Optimal working distance for the laser source. <br> 2. Alignment of the laser focus. <br> 3. Optimal purging gas flow. <br> 4. Heat sink for the nozzle <br> 5. Including recirculating coolant system. |

### 4.1.2.5 Impaired heat conduction:

Lack of fusion of the part to the substrate or sub-optimal support structure can lead to impaired heat conduction and erratic hot spots in the part. This might lead to geometric deformities and non-homogeneous microstructure. Buckling of the part and separation of the part from substrate can impair the heat conduction and lead further deformation due to excess heat.

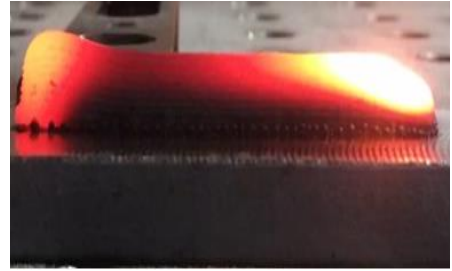

*Figure 27: Heat conduction is interrupted by lack of fusion defect at the start of the build [24].*

| CAUSE | MANIFESTATION | SOLUTION |
|---|---|---|
| 1. Lack of fusion of the component to the substrate [24]. 2. Absence of enough support structure to transfer heat. | 1. Erratic hot spots [24]. 2. Geometric deformities like bucking of the component. 3. Increased thermal stress. | 1. Use of more support structures to release the heat into substrate. 2. Use of higher power and lower scan speed |

### 4.1.2.6 Humidity and Impurity:

Humidity in the system can cause oxidation of the part during the build and cause contamination of the melt pool. Impurities present in the stock material and the system can be delivered to the melt pool and degrade the material properties. Impurities accrue in the system over prolonged use. Impurities may exist in the system due to improper handling, leak of the machine fluids, or by the artifacts of the build. Humidity can be removed by vacuuming the system before operation and proper handling can avoid cross-contamination of the system and the stock powder.

| CAUSE | MANIFESTATION | SOLUTION |
|---|---|---|
| 1. Water content in the system. 2. Vaporization of lubricant. 3. Artifacts of the build and precursor materials | 1. Oxidation/Dis-coloration of the component. 2. Inclusion of foreign particles causes contamination of weld pool and impairs fusion. | 1. Use of a pump to vacuum out chamber before purging with inert gas. |

### 4.1.2.7 Overheating of laser head:

Laser head may overheat due to heat radiation of the melt pool. Optical reflection of the laser beam from the melt pool contributes to the overheating.

| CAUSE | MANIFESTATION | SOLUTION |
|---|---|---|
| 1. Reflected laser from the deposition surface [50]. 2. Radiated heat from the melt pool and bulk of the part. | 1. Change in deposition characteristic. | 1. Installing a closed loop system for the chiller system. 2. Placing the laser head away from the nozzle assembly. |

### 4.1.2.8 Faulty powder stream patterns:

Powder stream may be affected by the flowability of the stock material. High scan speed can affect the inertia of the powder stream which may distribute the powder throughout the LENS chamber.

| CAUSE | MANIFESTATION | SOLUTION |
| --- | --- | --- |
| 1. Effect of flowability of powder to feed into the carrier gas stream [94]. <br> 2. The inertia of powder stream. <br> 3. The enhanced convection of powder in gas streams. <br> 4. Vapor stream. <br> 5. Turbulence due to high-speed gas stream. | 1. Reflectivity/absorptivity of the metal powder is affected and hence the deposition characteristics [95]. <br> 2. Reflectivity losses and spatter. <br> 3. Electrostatic repulsion. | 1. Systems must consider the effective feed rate of the feedstock, as appropriate amounts of deposit material must be delivered [94]. |

### 4.1.2.9 Flowability:

The flowability of the powder is dependent on particle size, shape and distribution. Large variation in shape and size distribution can lead to stochastic deposition and density.

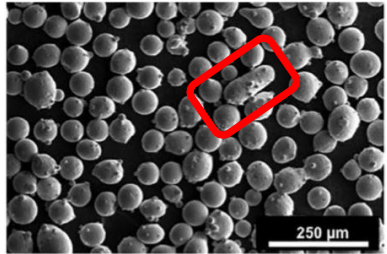

*Figure 28: Coalesced and out of shape powder particles.*

| CAUSE | MANIFESTATION | SOLUTION |
| --- | --- | --- |
| 1. Atomization defect [96] <br> 2. Improper handling of powder [96] <br> 3. Coalescing of powder particles <br> 4. Powder shape and size | Inconsistent powder flow | Ensure correct and uniform powder size and nozzle aperture and quality control. |

### 4.2 Controller Errors:

These errors occur due to erroneous sensor readings or inaccurate feedback systems. Out of calibration systems contribute to defects but may go unnoticed. These are difficult to diagnose and trace back. Intermittent maintenance, calibration and monitoring can be instrumental in averting these errors.

### 4.2.1 Laser Power:

Non-optimal laser power due to controller error can lead to melt pool anomalies and alter the deposition dynamics. An out of focus laser leads to inconsistent energy density and deposition characteristics.

| CAUSE | MANIFESTATION | SOLUTION |
|---|---|---|
| Offset error which changes the focal plane and beam size and hence the energy density | Undesired melt pool temperature which affects the deposition | Feedback system to check the focal plane. |

### 4.2.2 Erroneous reading of chamber pressure and gas fraction:

The erroneous reading of gas pressure in the chamber can lead to maladjustments by operator that may lead to various defects such as oxidation and porosity. Chamber pressure issues arise from leaks in the chamber or the supply lines.

| CAUSE | MANIFESTATION | SOLUTION |
|---|---|---|
| 1. Gauge error.<br>2. Leakage in the system. | 1. Inclusions of pores and cavities.<br>2. Oxidation | Multiple gauges to substantiate the pressure reading. |

### 4.2.3 Interpolation error:

Interpolation error of the deposition bed or the laser head results from an out-of-calibration system which can affect the part geometry. Monitoring of machine parameters and sensors data can help to ensure correct interpolation.

| CAUSE | MANIFESTATION | SOLUTION |
|---|---|---|
| Erroneous calibration of the drives. | Can greatly affect the dimension and deposition characteristic. | Intermittent calibration of the drives. |

### 4.2.4 Latency and Backlash:

The time delay in the communication can lead to error in interpolation and dwell time contributing to out-of-geometry prints. When too many process variables are communicated at once the communication bus can be busy and lead to latency.

| CAUSE | MANIFESTATION | SOLUTION |
|---|---|---|
| Bandwidth mismatch | Latency affects the switching time for the laser which directly effects the dwell time and run-offs. | Synchronous bandwidth for communication channel and laser pulse. |

### 4.2.5 Faulty optical scales and process sensors:

Sensor calibration is a necessary component of LENS maintenance. If this does not occur, the probability of faulty optical scales and process sensor readings grows.

| CAUSE | MANIFESTATION | SOLUTION |
|---|---|---|
| 1. Environmental states like temperature, pressure, vibration can attenuate the sensor readings.<br>2. Out of calibration systems. | 1. Material discontinuities commonly occur during processing.<br>2. Erroneous readings.<br>3. Offset values<br>4. Out of shape production. | 1. Calibration of sensors at different operational states.<br>2. Using channels that offer least attenuation.<br>3. Using a feedback loop can avert the error to some extent. |

### 4.2.6 Uncontrolled Jerks:

Abrupt changes in direction may result in particles obtaining momentum to carry past the intended deposition point. This causes deposition issues. In sections of the part in which axis acceleration changes significantly, such as where the deposition path changes from a straight line to curve, machine vibration or shock may occur. Speed control and a change of acceleration suppress vibration and machine shock and their associated machining errors.

| CAUSE | MANIFESTATION | SOLUTION |
|---|---|---|
| 1. Axis acceleration changes significantly in sections where scan path changes from straight line to curve. 2. Voltage spikes. | 1. Deformed or offset geometrical features might appear on the part. | 1. Speed control and a change of acceleration suppress vibration and machine shock 2. Alarms can inform of issues. |

## 5. MATERIAL ISSUES:

These types of anomalies deal with material-based originators of problems, either in the powder or via contamination that may occur in relation to the precursor powder material. These arise by virtue of faults during the production of stock material and the environmental conditions. The material can also degrade due to improper handling before, during or after the process.

### 5.1 Initial porosity in powder particles:

The powder particles may contain pores that get created during its production by gas atomization. The carrying over of such pores into the microstructure of the manufactured part will reduce part quality. During formation of the powder, liquid metal interacts with the inert gas environment during atomization, leading to the trapping of significant levels of gas within the particles, irrespective of the alloy or the production technique. Shrinkage that occurs during solidification forms finer voids. A series of studies by Zhong et al discovered that plasma rotate electrode process (PREP) powder yields less porosity than gas atomized powder.

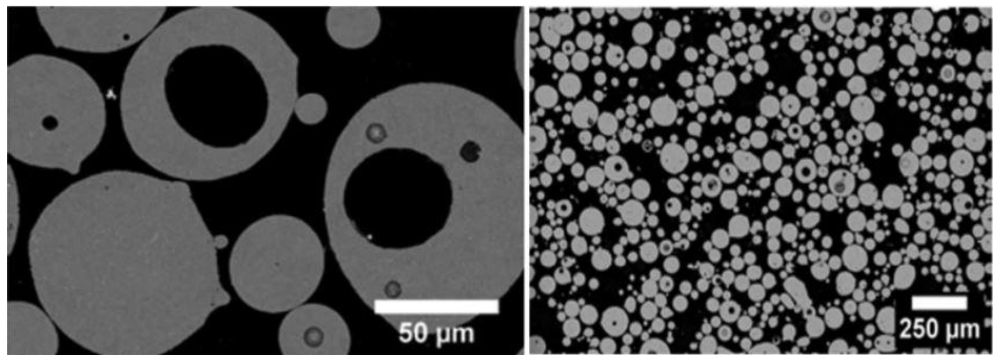

*Figure 29: Images of gas atomized MAR-M-247 powder, size range: 45-106 um containing porosity [97].*

| CAUSE | MANIFESTATION | SOLUTION |
|---|---|---|
| 1. Interaction of liquid metal with the inert environment, leading to the entrapment of gas within the particles [51]. 2. Solidification shrinkage. | When the powders are melted the gas entrapments coalesce and form pores and voids in the bulk [51]. | PREP yields less porosity than gas atomization approaches [19]. |

## 5.2 Surface satellites on powder particles:

The powder particles may be misshapen and contain surface satellites which alter the flowability and wettability of the powder. Studies show this aspect is directly linked to the equipment design and parameters used in powder atomization. These surface features impede continuous and homogeneous flowability when spread across a powder bed or when propelled by a carrier gas through the nozzle of a powder feeder. This variability in powder feeding can get in the way of part consolidation and promote the formation of porosity in AM structures.

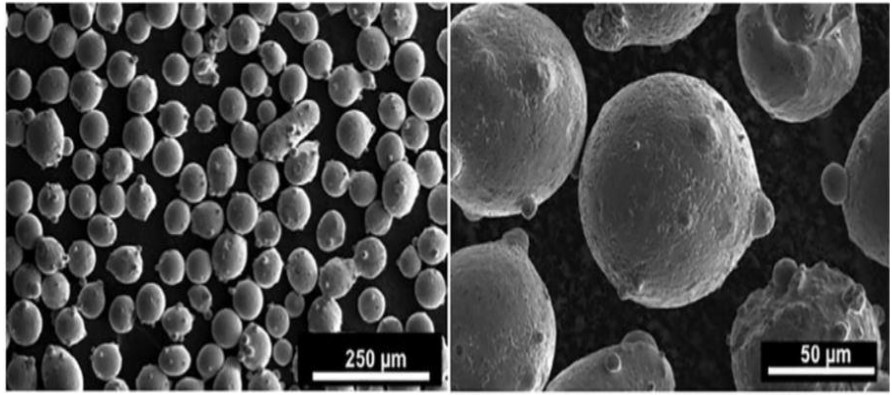

*Figure 30: SEM images of gas atomized MAR-M-247 powder, size range: 45–106 µm containing surface satellites [51].*

| CAUSE | MANIFESTATION | SOLUTION |
|---|---|---|
| Equipment design and parameters used in powder atomization [51]. | 1. Reduction of continuous and homogeneous flowability 2. Lack of part consolidation 3. Formation of porosity in finished structures. | 1. Optimized process parameters used in powder atomization 2. Quality assurance to eliminate defective powder. |

## 5.3 Difference in powder Size/Shape:

The powder size distribution affects the flowability of the particles and hence the deposition dynamics. The out of shape powder particles can block the nozzle and cause erratic deposition. Non-spherical particles should be eliminated before starting the deposition**.** Use of vibratory sieve can help to eliminate out of size particles prior to deposition.

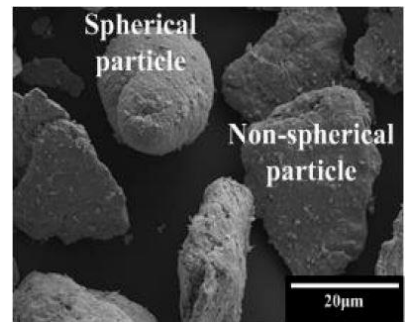

*Figure 31: Mixed powder with non-spherical particles [96].*

| CAUSE | MANIFESTATION | SOLUTION |
|---|---|---|
| Inappropriate production practice [96]. | The powder particles vary largely in their size and shapes [96]**.** | Use of vibratory sieve to eliminate out of size. |

## 5.4 Contamination of the powder:

The powder can contain contaminants which are introduced during the manufacturing process or handling by the operator. The powder can develop oxide contaminants through oxidation. For successful deposition, these should be eliminated to prevent contamination of melt pool. To reduce this defect, a high-performance vibratory sieve, where all contaminated particles are removed, should be used The qualified powder is collected in a container, ready for printing, and the oversize contamination is removed from the process without the loss of good powder.

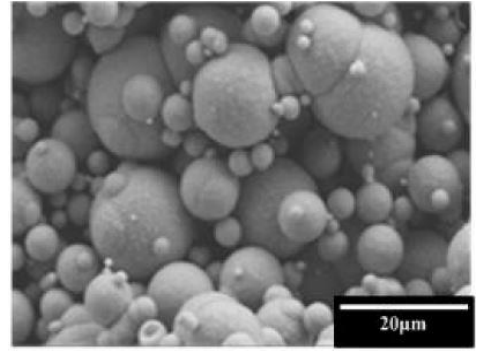

*Figure 32: Contaminated powder particles reproduced from [96].*

| CAUSE | MANIFESTATION | SOLUTION |
|---|---|---|
| 1. Inclusion of foreign particles [96]. 2. Oxidation 3. Humidity | 1. Difference in powder size. 2. Presence of Humidity. 3. Powder segmentation. | 1. High-performance vibratory sieve that removes contaminated particles. 2. Collected powder contamination is removed from the process [96]. |

## 5.5 Inclusions in Powder:

The powder can contain foreign particles which can stick on the part and remain. These can give rise to surface deformities and act as stress concentrators. These can exist as a manufacturing defect of the stock powder or due to inappropriate handling during fabrication and storage. Sieving can be an effective method to eliminate some of the foreign particles.

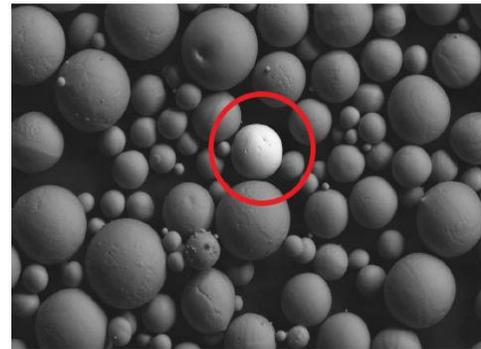

*Figure 33: Back-scattered SEM image of powder, highlighting the foreign particles [56].*

| CAUSE | MANIFESTATION | SOLUTION |
|---|---|---|
| Wear during powder production might lead to inclusion of foreign particles [56]. | Existence of foreign particles. | Pre-processing and sieving to eliminate bigger foreign particle [56]. |

## 5.6 Sphericity:

The powder particles can be out of sphericity due to atomization defect. The powder's sphericity can also change over time if not properly handled and stored. Micrograph analysis and sieving may reveal powder particles that are not within range of acceptable sphericity.

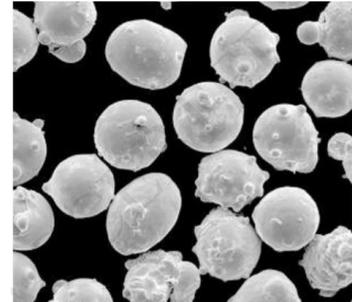

*Figure 34: SEM image of out of shape powders.*

| CAUSE | MANIFESTATION | SOLUTION |
|---|---|---|
| Atomization defects [96]. | Particles that are not within range of acceptable sphericity. | Sieving helps to remove out of size particles [96]. |

## 6. PRODUCTIVITY ISSUES:

Productivity issues create a toll on the cost of the job and machine down-time. This can be in the form of issues inherent to the process or design.

### 6.1 Non-optimal AM part design increases manufacturing time:

The process cycle time could be significant and cost-prohibitive if the part design is not optimized for AM, besides build failure.

| CAUSE | MANIFESTATION | SOLUTION |
|---|---|---|
| 1. Part design which is not suitable or designed for AM. 2. Non-commensurate process parameters. | 1. High machine running time. 2. Build failure. | 1. Optimized design for additive manufacturing can reduce the production cost. 2. Weight reduction. 3. Optimized process parameter as per the design. |

### 6.2 Need for Post-Processing:

Additional post-processing by conventional machining processes may be necessary to bring the AM part to the desired surface finish and dimensions. This increases the cycle time. Some AM parts may be too difficult to post-process via machining due to its inherent complexity and difficulty to fixture.

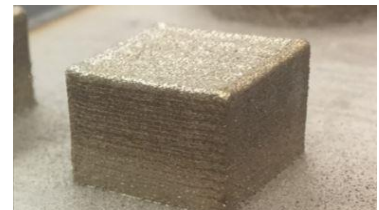

*Figure 35: Surface finish of the as-deposited part.*

| CAUSE | MANIFESTATION | SOLUTION |
|---|---|---|
| 1. Surface deformation. | 1. Poor surface finish. | 1. Production of near net shape. |
| 2. Fused powder on the surface. | 2. Deformed contour. | 2. Reduced heat input. |
| 3. Shrinkage that causes uneven surface | 3. Material overlay. | |
| 4. Excess material that may accumulate. | | |
| 5. Pores which are produced. | | |

## 6.3 Powder wastage:

Some powder may be left unused and unmelted here (~50%+ powder is not fused onto the part) which can be collected and reused for subsequent processes. In some cases, this is not viable due to concerns about contamination.

| CAUSE | MANIFESTATION | SOLUTION |
|---|---|---|
| Inefficient powder delivery. | Wasted powder sprayed within LENS chamber. | 1. Processes to purify the recovered powder. 2. Re-Homogenization. 3. Recycling. |

## 6.4 Poor Machinability:

Due to non-homogeneous microstructures created as parts are exposed to continuous thermal cycling, it may be difficult to machine the part. This adds to the part cycle time and the tooling cost. Due to rapid and non-uniform cooling rates in DED, the microstructures could vary significantly. Heat treatment of these components can alter some mechanical properties, and consequently improve machining productivity.

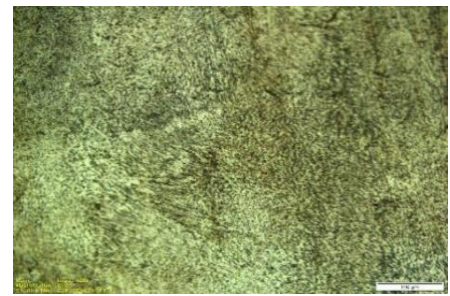

*Figure 36: Fine equiaxed structure of SS316LN produced by DED.*

| CAUSE | MANIFESTATION | SOLUTION |
|---|---|---|
| Rapid cooling process leading to less machinable microstructures. | Poor machinability in resulting components. | Heat treatment can change mechanical properties suitably, and thereby difficulty in machining. |

## 6.5 De-powdering and removal of support material:

After the process is accomplished the support structures and unfused powder must be removed. The time taken to remove them varies with component complexity and accessibility.

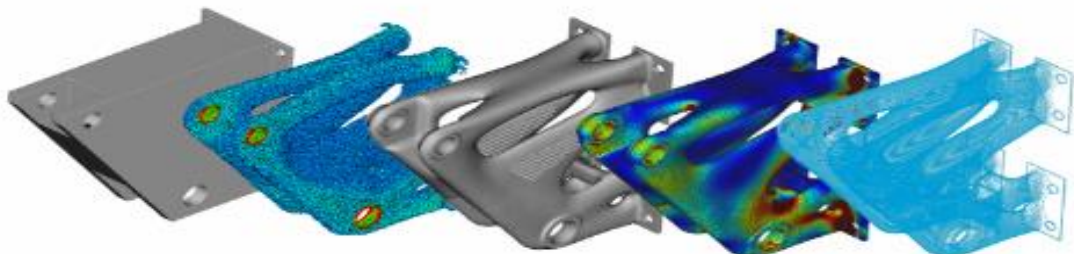

*Figure 37: Topology optimization to use minimal support structure.*

| CAUSE | MANIFESTATION | SOLUTION |
|---|---|---|
| 1. Support structures that are created increase post-processing time. 2. Support structures act as a heat sink. | Heat sink affects the melt pool dynamics and hence the structural integrity of the build. | 1. Optimized Design for AM. 2. Topology Optimization. |

## 6.6 Power Consumption:

If the design is not optimized for AM it can lead to excess power consumption which may cause the process to become cost prohibitive.

| CAUSE | MANIFESTATION | SOLUTION |
|---|---|---|
| 1. Over/Under powered laser. 2. Slow scan speed 3. Higher dwell time 4. Lower layer thickness | Higher machine running time. | 1. Optimize layer thickness 2. Lower dwell time. 3. Reduce unnecessary retraction. 4. Optimized scan strategy to reduce laser path [86]. 5. Fractal and offset strategies could attract more attention due to geometrical accuracy and reduced energy consumption [87]. |

## 7. SAFETY ISSUES:

Safety issues concern any occurrence that may cause injury and endanger the life of the operator or the lifecycle of the machine. These are mentioned with little focus in literature but deserve strong focus to help improve the validity of DED machines.

### 7.1 Crashing nozzle head into the stage due to programming:

The nozzles can crash into the part if the nozzle dimensions are not considered while programming the part. Certain geometries might need a different slicing strategy to avoid damage to the nozzle head.

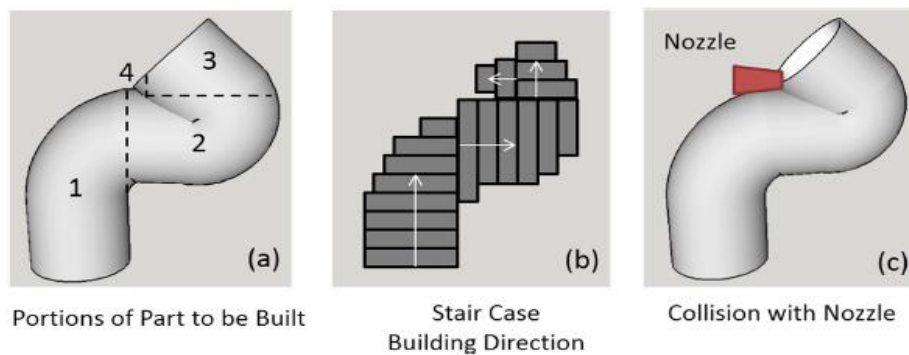

*Figure 38: Multi-axis processing methods using paralleled and non-paralleled layer approaches as reproduced by [46].*

| CAUSE | MANIFESTATION | SOLUTION |
|---|---|---|
| The nozzle head can possibly crash into the stage due to erroneous programming, offset or hardware faults. | Nozzle head can crash into the stage and cause damage that has to be fixed by a technician and puts the machine down for a considerable time [46]. | Meticulously planned part programming to avoid obstructions and possible collision points. |

### 7.2 Fire Hazard:

The high surface to volume ratio of powder used for developing the part makes it prone to fire and can catch fire if not maintained correctly, exposed to oxygen and a large amount of heat is introduced.

| CAUSE | MANIFESTATION | SOLUTION |
|---|---|---|
| Fires can start in the machine if improper process parameters are used with powder issues. | Smoke or fire can be seen. | Ensuring inert environment while handling powders. |

### 7.3 Powder Safety:

The particles can remain suspended in air and pose a health hazard, if inhaled. Exposure to airborne powder particles can cause lung damage to operators, especially when particles are inhaled on a regular basis. To avoid this, an operator should follow the Standard for the Prevention of Fire and Dust Explosion from manufacturing, processing, and handling of combustible particulate solids.

| CAUSE | MANIFESTATION | SOLUTION |
|---|---|---|
| Combustibility of AM powder (High surface to volume ratio) | Health hazard due to particle inhalation | Follow standard for the Prevention of Fire and Dust Explosion |

### 7.4 Exposure to dangerous levels of invisible radiation:

The DED system works on a class I laser which can cause severe injuries if an operator is exposed directly.

| CAUSE | MANIFESTATION | SOLUTION |
|---|---|---|
| Class I lasers can lead to severe injuries if proper safety measures are not taken. | 1. Irritation<br>2. Itching<br>3. Burning Sensation | 1. CDRH-certified viewing window.<br>2. Use of interlocks. |

### 8. REPAIR-SPECIFIC ISSUES

The repair of damaged components is an important aspect of DED for both economic and manufacturing reasons [98]. Repair can be an option when remaking a component is cost prohibitive. Re-creation of the component through LENS may then introduce performance inconsistencies if similar parts were made through traditional means. There are various issues that accompany the use of LENS as a repair device. To best mitigate these effects, operators should be aware of the following issues.

### 8.1 Excess delivery of repair material:

The machine may overdeliver the amount of material at a location to fix an issue. Another issue may occur when an operator or the programming dispenses powder to a region that was intended to be pure or kept at a certain composition. This can lead to a change in composition and material properties if enough powder is incorporated into the region. Secondary phases may develop if continuous addition occurs. Proper monitoring of where molten powder is delivered can help mitigate issue. This may include sensor systems that alert the user that unintended delivery to a region may be occurring.

| CAUSE | MANIFESTATION | SOLUTION |
|---|---|---|
| Operator or programming error | 1. Change in composition and material properties | 1. Proper monitoring of powder delivery |

### 8.2 Mismatch in material properties for repair deposits:

A step change in material properties may develop due to the nature of repair deposits. These are typically single beads or tracks that act differently from a layer by layer build. Linear heat input plays a role in the measured difference in properties. This may be due to higher cooling rates in low energy repair that may create different phases and properties. Nassar et al. utilized a closed-loop control based on previous material temperature before deposition which reduced the incidence of the hardness gradient they encountered [99]. Kistler et al. found that hardness decreased as the temperature of the substrate was increased prior to operation. Interlayer dwell time was shown to have the second highest influence on deposit hardness [98]. Strategies from Section #3.3.2.2 can be used.

| CAUSE | MANIFESTATION | SOLUTION |
|---|---|---|
| 1. Repair deposits causing change in properties [16, 100]. <br> 2. Linear heat input [101]. <br> 3. Higher cooling rates in low energy repair. | Deposits may have higher or lower properties than the material location being targeted for repair [102]. | 1. Closed-loop control based on temperature to reduce property gradients <br> 2. Increase temperature of substrate <br> 3. Optimized interlayer dwell time |

### 8.3 Repair strategy may cause distortion in parts:

Heating of a part with individual deposits may cause distortion of the structure from the intended design. The repair strategy may cause parts to become misshapen when a new heat affected zone is created from repair deposits made. Custom repair solutions that are a combination of modeling and experimental validation should repair components the best. Controlling cooling rates before and after deposition can help mitigate deformation.

| CAUSE | MANIFESTATION | SOLUTION |
|---|---|---|
| 1. Repair deposits causing new heat affected zones 2. Skewed or concentrated heat area | 1. Parts may undergo structural changes when introduced to the high temperature deposits. 2. Warping | 1. Optimized processes that can more efficiently refill damaged parts [103]. 2. Modeling and experimental validation 3. Controlling cooling rates |

**8.4 Repair strategy may cause residual stress formation:**
Depositing singular beads or lines of material on the surface of the part to repair a part may cause residual stress to form within the bulk. Strategies can be employed from Sections 3.2.2.2., 4.1.1.4, and 8.3.

| CAUSE | MANIFESTATION | SOLUTION |
|---|---|---|
| Deposit of singular beads or lines of material shot on the surface of part. | Parts may develop residual stresses within the bulk due to stresses imposed by the new deposits. | 1. Optimized processes that can more efficiently refill damaged parts [103]. 2. Modeling and experimental validation 3. Controlling cooling rates |

# 9. COMPOSITIONAL ISSUES

**9.1 Cracks and delamination due to the deposition of different elemental compositions:**
Cracks and delamination may be manifest between two material types joined. Partial melting of powders may result in lack of fusion defects [104]. Factors such as delamination or cracking may become prominent from large material property differences at the interface [50]. Deposition of coatings with a very high TiC content directly onto a Ti substrate was shown to cause cracking at the interface between substrate and coating. Cracks and delamination may occur after a certain amount of new material layers are deposited [105]. Using optimized process parameters and proper compositions, crack-free deposits can be produced [15]. Material property difference issues can be mitigated through more gradual stepwise changes in space [50]. Strategies from Sections #3.2.2.1-3.2.2.3 and 3.2.2.7-3.2.2.8 can be used.

| CAUSE | MANIFESTATION | SOLUTION |
|---|---|---|
| 1. Partial melting of powders 2. Significant temperature difference between the initially formed layers and newly formed layers. 3. Large material property differences at interfaces | 1. Cracking at the interface between substrate and coating 2. Cracks and delamination | 1. Optimized process parameters and proper compositions 2. Gradual stepwise changes in space |

## 9.2 Partial melting of powders due inconsideration of different powder properties:

Without adjustment for the addition of another material, powders may not completely melt due to improper process parameters. Different powders possess different densities and material properties. Laser absorption coefficient, melting point, concentration, and enthalpy of mixing have a substantial effect on chemical consistency the necessary laser energy. Blends of elemental or alloy powders can be prepared that are chemically and microstructurally identical to the pre-alloyed powders [15]. Lower melting point materials should be delivered before the movement of the laser. Higher melting point materials should be delivered behind the movement of the laser [106].

| CAUSE | MANIFESTATION | SOLUTION |
| --- | --- | --- |
| 1. Operator ignorance of different powder properties 2. Non-optimal process parameters | Unmelted or partially melted powder may be observed in the first layers of the interface from one material to the next [15]. | 1. Proper blending of powders 2. Optimized powder delivery strategy |

## 9.3 Intermetallics and secondary phases from the combination of two dissimilar materials:

The introduction of another metal powder to a single element or alloy structure may cause the development of intermetallics and secondary phases [107]. Diffusion between two materials may cause intermetallics to form between the two materials. Intermetallics form at the interphase or junction between graded layers which reduce the structural integrity of the material. The design of a multi-material part should include a consideration for the tolerance to elemental distribution [55]. Intermediate layers between sharp gradients can help mitigate the formation of intermetallic phases. Consideration of secondary phases that may form should be factored into the development of gradients to better accommodate the build [108]. Strategies from Sections #3.3.2.1 and 3.3.2.2 can be used.

| CAUSE | MANIFESTATION | SOLUTION |
| --- | --- | --- |
| 1. Introducing a certain amount of metal powder to a system to cause microstructure changes. 2. Diffusion between materials | Intermetallics form at interphase between graded layers | 1. Design considerations for distribution of elements 2. Intermediate layers |

## 9.4 Powder delivery without consideration for density:

When developing compositional gradients through LENS, the powder density, size, shape, and surface morphology of each powder used must be considered so proper deposition can occur. An operator who does not consider the density of powder when performing calculations to determine the percentage for each gradient step will deliver the wrong amount of powder and a corresponding error in composition. An unintended composition results from the incorrect amount of powder being delivered. The quality of builds can be difficult to control due to segregation effects in the powder blends [106]. To mitigate these issues, an operator should perform a flowability test to determine the delivery rate of the powder with respect to the powder density. Elemental powder blends with premixed powders can be placed in the feeder [106]. A dual strategy of elemental powder blend and delivery can be employed to lead to the right structure.

| CAUSE | MANIFESTATION | SOLUTION |
|---|---|---|
| Negligence of powder density when performing powder deposition | 1. Wrong composition for part 2. Quality of build is affected. | 1. Performing calculations with the density of powders measured against the powder delivery rate. 2. Premixed powder blends |

### 9.5 Lack of bonding between two dissimilar materials:

Lack of bonding may occur between compositionally graded layers as a structure is built. Residual stresses and limited miscibility of two alloys may cause a lack of bonding [18]. The formation of brittle intermetallics between layers can cause a lack of bonding. Thermal expansion can occur between two different materials that cause structural integrity issues [97]. Low bond strength can lead to failures like cracking or stress formation at the interface. Large thermal fluctuations can cause some structures to come out with poor structural integrity. An interfacial bond is an important consideration to maintain the performance of the final part [18].

| CAUSE | MANIFESTATION | SOLUTION |
|---|---|---|
| 1. Residual stresses 2. Limited miscibility 3. Formation of brittle intermetallics 4. Thermal expansion between different materials | 1. Cracking and stress formation at the interface 2. Poor structural integrity | Considerations of interfacial bonds |

## 10. SOME TECHNOLOGICAL AND SCIENTIFIC GAPS AND R&D NEEDS:

Despite almost two decades of research on DED processes, there is a lack of thorough understanding of and control over the various defects, their pathways and root causes in the form of process faults and the precursor material issues, as well as the implications of these anomalies on safety and productivity. These gaps hamper a wide industrial application of this versatile process. The present effort is a first step towards delineating these complex process-product-performance relationships inherent to the DED process. This work was based on combining the evidence (authors' experimental observations) with existing literature. The framework of this paper is to categorize and elucidate various anomalies, and we anticipate that this would spur additional work to supplement this knowledge and extended to other novel materials.

A similarly detailed sister compendium regarding the process parameter ranges for various material classes would be an important contribution to the state of the art. While a large amount of results would be required to capture the behavior of various materials, these efforts would represent a significant contribution to the current knowledge. A collection of repeatable procedures would reduce the amount of information-search and experimentation to explore and optimize the process. Instead, research work could be focused on scientific pursuits of the process mechanisms, as well as creative pursuits surrounding the application and advancement of the process. These efforts can address the present need to achieve repeatability and control of the material flow, the thermal history, as well as the machine motion to realize various geometric, morphological, and microstructural features. While experienced operators have developed mostly experiential guidelines to address various anomalies in DED, a wide industrial adoption of this process demands an in-depth understanding of the process. A compendium of best practices improves process success for both known and yet to be developed structures. The adoption of standard operating procedures can help facilitate process recipes that can be extrapolated to future DED-based development.

Due to the complex multiphysics of the DED process, as well as the diverse relationship of the defects with the process and material anomalies therein, a DED process demands extensive monitoring to precisely control the build. With recent advances in sensing and imaging technologies, effective and economical smart sensing solutions would enhance quality assurance and integrity of the DED process [109].

Additionally, very little exploration has been made about how DED-based technologies could affect the industrial ecosystem. With the speed of technology improvement, it is plausible that DED could become table stakes for a narrow industrial sector to for metal manufacturing. While incremental experimentation could bring great value to unlocking these understandings there are still various aspects that have not been explored.

## 11. CONCLUSION AND FUTURE WORK:

The present work has provided an extensive review of anomalies in a DED process as well as a framework for categorizing process anomalies under seven major categories that can be applied and referenced easily. Despite efforts on various aspects of AM research, a refined unified understanding of the process issues does not exist heretofore. Extensive reviews offer an abundance of information, but the breadth provided by these compendiums also introduces unnecessary complexity to the study of these systems. Studies of AM process anomalies have been pored over in specific contexts, but no attempt has been made to catalogue the types of anomalies and offer a framework for analysis henceforth.

In this review, the prevalent issues are detailed along with either multiple solution sets or proposed solutions that may be analogous to general manufacturing malfeasance. Development of new materials with enhanced capabilities for AM processes can make the process more robust and efficient. Such development requires a fair understanding of material and process capabilities, as well as the material-process-property relationship.

The DED process as well as DED-processed parts have various issues in terms of part quality, process environment and efficiency due to the nature of interacting internal parts and conditions. These issues are further complicated by the dynamics of the process parameters used. This work has enumerated the most common defects and anomalies for DED with plausible causes. Feasible solutions to mitigate these issues are delineated in a pointwise fashion. A causal relationship is established between various process parameters and their contribution to defects. To eliminate or minimize a defect the necessary parameters can be observed through the heat map. Further monitoring and optimization can assist with producing better build quality. There cannot be a single-handed approach to mitigate all defects due to inherent complexity of the process.

To maximize the utilization and bring the process on par with highly reliable conventional manufacturing practices, the productivity issues need to be addressed to improve the manufacturing feasibility and reduce the associated costs. Safety concerns, which pertain to catastrophic errors that may cause damage to the operator or the machine, are paramount to worker safety and continued adoption of LENS-DED machines. Contamination issues that arise from inherent or additional equipment within DED must be considered to improve the safety of the operator and to avert detrimental effect on the environment.

Further work on the safety of laser systems in a commercial or home setting could bring a focus on the feasibility of DED systems and other laser powered AM machines for widespread use. Until the systems that reduce the amount of radiation and other unpleasant side effects of DED operation are optimized, laser-based AM devices will remain tethered to laboratories. Furthermore, an ethical study of the availability of designs based on metal-based prints should be undertaken. There are numerous novel capabilities afforded by these machines. With this great power, a renewed sense of responsibility is necessary. Thus, awareness of the possible ethical issues must be explored to further feasible use and adoption of these machines for the future.